\documentclass[aip,jcp,preprint,showpacs,amsmath,amssymb]{revtex4-1}

\usepackage{graphicx}
\usepackage{dcolumn}
\usepackage{bm}
\usepackage{ae}
\usepackage{aecompl}
\usepackage{amsmath}
\usepackage{amssymb}
\usepackage{epsfig}

\newcommand{\prom}[2]{\left \langle #1 \right \rangle_{#2}}
\newcommand{\qvec}{\mathbf{q}}
\newcommand{\Eq}[1]{Eq.~(\ref{eq:#1})}

\newcommand{\sech}{\rm sech}

\newcommand{\rvec}[2]{\mathbf{r}_{#1}^{#2}}
\newcommand{\rpar}{\mathbf{x}}
\newcommand{\film}{\mathcal{L}}
\newcommand{\funcL}{\film}
\newcommand{\ellav}{\ell}
\newcommand{\rhoi}{\rho_{\pi}}

\newcommand{\hver}{h}
\newcommand{\hverpi}{h_{\pi}}
\newcommand{\hper}{s}
\newcommand{\dhs}{\sigma}

\newcommand{\alfa}{\alpha}
\newcommand{\exilon}{\epsilon}
\newcommand{\strength}{u}
\newcommand{\mi}{m_{\pi}}

\begin{document}

\title{
Capillary Wave Theory of Adsorbed Liquid Films and the Structure of the
Liquid-Vapor Interface
}

\author{Luis G. MacDowell}
\affiliation{Departamento de Qu\'{\i}mica F\'{\i}sica, Facultad de Ciencias
Qu\'{\i}micas, Universidad Complutense, Madrid, 28040, Spain}

\begin{abstract}
In this paper we try to work out in detail the implications of a microscopic
theory for capillary waves under the assumption that the density is given 
 along lines normal to the interface. 
Within this approximation, which may be justified in terms of symmetry
arguments, the Fisk-Widom scaling of the density profile holds for frozen
realizations   of the interface profile.
Upon thermal averaging of capillary wave fluctuations, the resulting density
profile yields results consistent with renormalization group calculations in the one loop approximation. 
The thermal average over capillary waves may be expressed in terms of a modified convolution approximation
where normals to the interface are Gaussian distributed.  In the absence of an external field
we show that the phenomenological density profile applied into the square 
gradient free energy functional recovers the capillary wave Hamiltonian  exactly. 
We extend the theory  to the case of liquid films adsorbed on a
substrate. For systems with short range forces, we recover an effective interface
Hamiltonian with a film height dependent surface tension that stems from the distortion
of the liquid-vapor interface by the substrate, in agreement with the
Fisher-Jin theory of short range wetting. In the presence of  long range
interactions, the surface tension picks up an explicit
dependence on the external field and recovers the wave-vector dependent logarithmic 
contribution observed by Napiorkowski and Dietrich. 
Using an error function for the intrinsic density profile, we obtain
closed expressions for the surface tension and the interface width.
We show the external field contribution to the surface tension may be given
in terms of the film's disjoining pressure. From literature values of
the Hamaker constant, it is found that the fluid-substrate forces may
be able to double the surface tension for films in the nanometer range.
The film height dependence of the surface tension described here is
in full agreement with results of the capillary wave spectrum obtained
recently in computer simulations, 
and the predicted translation mode of surface fluctuations 
reproduces to linear order in field strength
the exact solution of the density correlation function
for the Landau-Ginzburg-Wilson Hamiltonian in an external field.
\end{abstract}

\maketitle

\section{Introduction}
The structure of  the liquid-vapor interface and the corresponding
capillary wave fluctuations continue to receive a great deal of attention after
many decades of research.\cite{hofling15,parry16,chacon16}  Under the mean field approximation,
the statistical mechanics of interfaces is most conveniently expressed
in terms of Density Functional Theory.\cite{evans79,evans92} This approach provides
an {\em intrinsic} density profile, which only depends on molecular details
of the fluid under study.
A wide-reaching implication is the Fisk-Widom scaling hypothesis,
which suggests that close to the critical point the density profile 
becomes universal.\cite{fisk69,rowlinson82b} However, already within the mean field
approximation, a more detailed study of density correlations indicates that liquid-vapor 
interfaces exhibit a long wavelength instability, whence,  divergent fluctuations in the 
thermodynamic limit (however far from the bulk critical
point).\cite{zittartz67,jasnow78,evans15,parry16} This situation
implies that an accurate description of the interface must be carried out
within the framework of renormalization group theory.  Explicit calculations for
simple models show that the correct averaged density includes the mean field
intrinsic density profile as leading order contribution. However,
to second order a new term  appears which
does not conform to the Fisk-Widom scaling, but is rather, extrinsic, i.e.,
it depends also on the system size,
at least on scales smaller than the  {\em parallel correlation length}, $\xi_{\parallel}$ that is of macroscopic
range for a fluid interface under
gravity.\cite{jasnow78,abraham81,jasnow84,kopf08,delfino12}

A far more intuitive approach to the study of interface fluctuations may be
achieved in terms of capillary wave theory.\cite{buff65,weeks77,bedeaux85b} Here, one
assumes that surface fluctuations may be singled out from bulk fluctuations
by performing a pre-average on the length-scale of the bulk correlation
length.\cite{weeks77,huse85} The properties of the undulated
film profile that results may be then studied analytically, and it is found that the origin
of the diverging structure factor may be traced to capillary wave fluctuations of the
interface.\cite{jasnow78,evans15,parry16}  The thermal average of such fluctuations provides
an extrinsic interface width that is proportional to
$\ln\xi_{\parallel}$, in agreement with renormalization group
theory and exact calculations.\cite{jasnow78,abraham81,fisher82,kopf08}

X-ray scattering experiments as well as computer simulations have confirmed
the predictions of renormalization group and capillary wave theories, but
also indicate that the divergence of fluctuations is in practice a minor
concern for typical macroscopic samples.\cite{braslau88,ocko94,doerr99,benjamin92,mueller96b,lacasse98,vink05} 

Be as it may, the presence of an extrinsic interface width indicates an
important conceptual limitation of the usual mean field approach. For this
reason, efforts have been devoted to incorporate the parallel 
interface fluctuations within density functional theory and to
account for capillary wave fluctuations at the microscopic
level.\cite{davis77,napiorkowski93,robledo97,mecke99b,stecki01,blokhuis09} 
Particularly, recent studies have emphasized the need to account
for the interface curvature, and indicate that it is possible to
recover an effective capillary wave Hamiltonian from fully microscopic
functionals, provided one considers  an extended wave-vector dependent
surface tension.\cite{mecke99b,blokhuis09} Unfortunately, it has also been argued convincingly that it is not
possible to determine unambiguously these wave-vector dependent corrections
to the surface tension from x-ray scattering
experiments.\cite{paulus08,pershan12,parry16} The reason is
that surface and bulk fluctuations entangle at the large wave-vectors
that would be required to measure such corrections. Whence, the only
way to study interface fluctuations at small length-scales is 
adopting an arbitrary but consistent prescription for the interface
location and measuring its fluctuations by means of computer
simulations.\cite{chacon03,chacon05,tarazona07}

An apparently unrelated issue is the study of
{\em short range wetting}, i.e., the transition that takes place
when the only driving force to wetting is a very short range attractive
interaction of the fluid to the substrate.\cite{parry12,bryk13} In this limit, as the film
thickens the liquid-vapor interface fluctuations become large, and are
akin to the usual capillary wave fluctuations of a free interface. 
Theoretical studies on this topic indicate  that the
substrate distorts the liquid-vapor profile,\cite{jin93,parry06} and therefore conveys 
a film height dependence to the surface tension (also known as position dependent
stiffness)  of which there are currently strong indications from computer
simulations.\cite{fernandez12,fernandez15} 

Recently, we studied the interface fluctuations of an adsorbed film
in the presence of a long range external
field.\cite{macdowell13,macdowell14,benet14b} In this case, the liquid-vapor
interface feels the substrate directly via the long range forces, rather than
indirectly, via weak substrate-fluid correlations.   As a result, the
surface tension picks up a strong  film height dependence, which
increases with the intensity and range of the external field.\cite{bernardino09} Indications
of this effect observed already some time ago\cite{werner99} have been confirmed by
a number of recent simulations, which show that the film height dependence
may be related to the film's disjoining pressure.\cite{macdowell13,macdowell14,benet14b} 

Already a while ago, Davis suggested  that
a microscopic explanation of capillary waves may be achieved by assuming
the density is given in terms of the perpendicular distance to the interface
position.\cite{davis77} This idea, which looks quite intuitive 
and may
be justified from microscopic free energy
functionals,\cite{diehl80,kawasaki82,huse85}    
 has been henceforth
explored in depth.\cite{mecke99b,stecki01,blokhuis09} 
However, it appears that some of its implications
may have been overlooked.  In a recent paper, we showed that in
fact it is able to explain accurately the interface fluctuations in the
presence of long range external fields, and particularly, the relation
of the surface tension with the disjoining pressure.\cite{benet14b} A more
direct test of this hypothesis may be obtained from calculations
of density profiles of absorbed films.\cite{nold14,hughes15,nold15} Particularly,
accurate density functional
calculations of the density profile in the vicinity of the three phase
contact line (i.e., the rim of sessile droplets) by Nold et al. have confirmed that the
hypothesis is valid for adsorbed films even a few molecular diameters away from the 
substrate.\cite{nold16}

In this paper we try to work out in detail the implications of a microscopic
theory for capillary waves under the assumption that the density is given 
 along lines normal to the interface.\cite{davis77} Our study
provides interface Hamiltonians for adsorbed films in a variety of systems,
and shows that the corrections to the classical capillary wave spectrum
are of the same order as the surface tension. 
Whereas it seems difficult to disentangle the signature of such corrections
in surface scattering experiments, they seem to be in full agreement
with  recent computer simulations.\cite{macdowell13,macdowell14,benet14b}
Interestingly, our study also
sheds some light on the nature of the liquid-vapor interface
in the absence of external fields and allows us to reconcile the Fisk-Widom
scaling hypothesis with capillary wave theory.

In the next section  we make some general remarks that motivate the
phenomenological approach that is adopted here. We then formalize
the approximation and discuss its implications as regards
the structure of the density profile (Sec. III). The study follows with the formulation of
effective interface Hamiltonians for a variety of fluid-fluid and
fluid-substrate interactions (Sec. IV), which are then applied for a simple
intrinsic density profile with the shape of an error function (Sec V).
Finally, in section \ref{sec:exact} we compare our predictions
with exact solutions for the Landau-Ginzburg-Wilson Hamiltonian.
Our
findings are summarized in the conclusion.

\section{Preliminary Definitions}

\label{preliminary}

\begin{figure}
\includegraphics[width=0.7\textwidth]{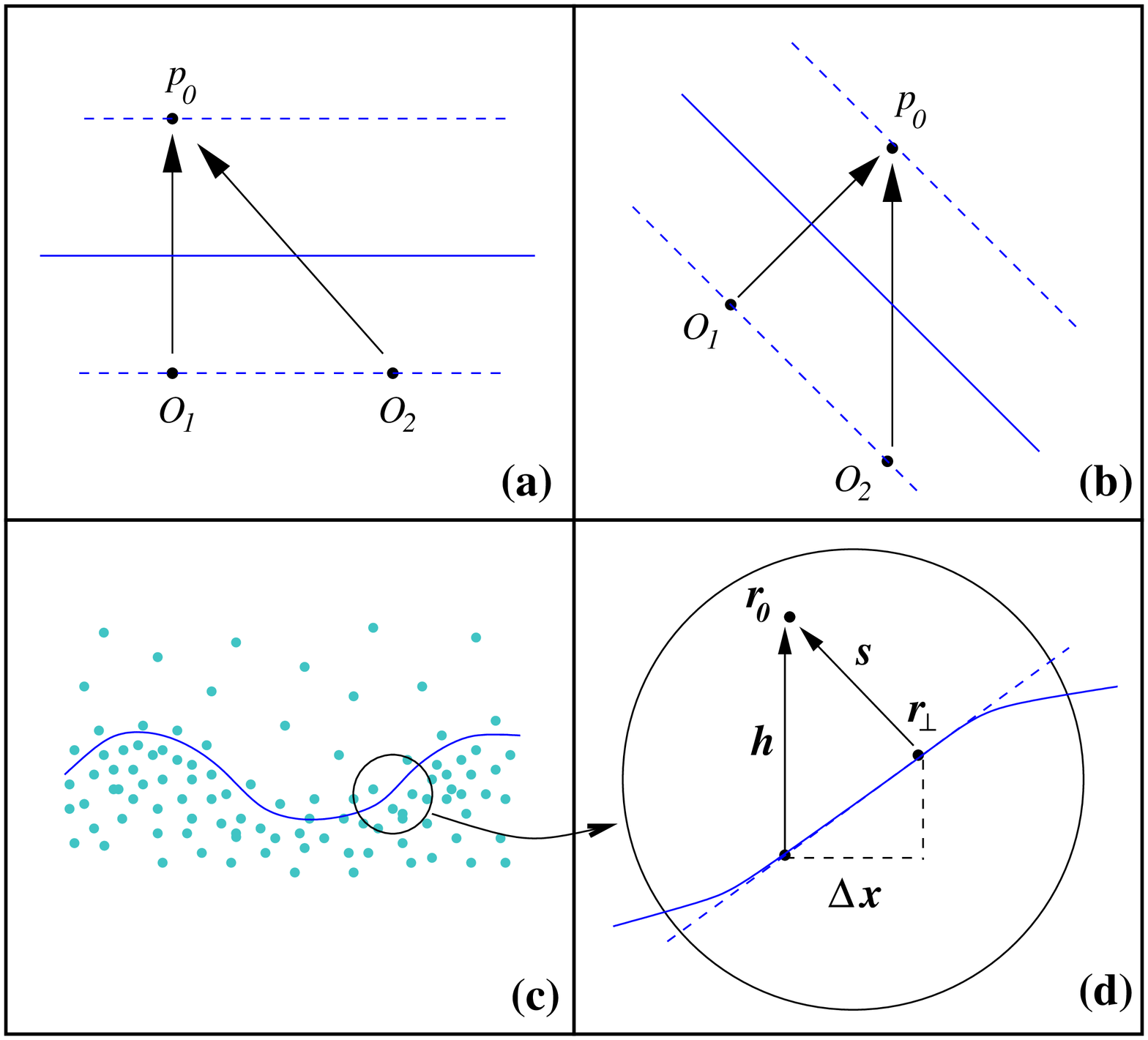}
\caption{\label{fig:sketch}  Sketch of simple model interfaces. (a) For
a flat interface profile (full line), choosing a reference frame $O_1$
perpendicular to the interface, one finds  isodensity lines (dashed)
are given by the vertical distance  $z-\ellav$. (b) For an arbitrary reference
frame, $O_2$, the interface appears tilted, and the isodensity lines depend simultaneously
on $\rpar$ and $z$. (c) At a microscopic scale a smooth film profile
may be defined after averaging at the scale of the correlation length.
(d) At a scale (circle) that is smaller than the curvature
of $\film$ but larger than the correlation length, the interface
looks flat but tilted. The perpendicular distance from the film to
a point $\rvec{0}{}=(\rpar_0,z_0)$ may be determined approximately as a local function of 
$\rpar_0$.
}
\end{figure}

\subsection{Symmetry}

Consider an atomic fluid in a state of vapor-liquid coexistence. 
A configuration of the system may be specified
in terms of the instantaneous density $\hat{\rho}(\rvec{}{})$, as dictated
by the set of atomic coordinates of the fluid. This density field is
highly discontinuous, but a related continuous density $\rho(\rvec{}{})$ may 
be determined as a thermal average of $\hat{\rho}(\rvec{}{})$ on the scale
of the correlation length. 
Having $\rho(\rvec{}{})$ at hand, it is
possible to define an interface  as the loci of points with
a prescribed density laying between bulk liquid and vapor densities.
Alternatively, from a given configuration, the interface location
may be specified using a smooth density operator with width equal to
the bulk correlation length,\cite{willard10} or by a suitable percoleation
algorithm.\cite{chacon05,tarazona07} 
In either case, a hypothetical situation may be envisaged where the 
thermal fluctuations of the interface position have been supressed.  
A point
in space, $\rvec{}{}$, may be given in terms of $\rpar$ and $z$, where the
latter is a direction perpendicular to the interface, and $\rpar$ is a
vector perpendicular to $z$. 
Choosing a suitable dividing surface, the corresponding  planar film profile,
say $\pi$, located at position $z=\ellav$,
is completely flat and devoid of any roughness at all length scales beyond the
bulk correlation length, as sketched in Fig.\ref{fig:sketch}.a. 
The density, $\rho(\rvec{}{})$, which will generally depend on $\rvec{}{}$ 
in this case is a single function of the distance  $\hper$ 
away from the interface. This serves to define an intrinsic density profile
$\rhoi(\hper)$, which is defined  here as the mean field density profile obtained
from the underlying microscopic free energy functional.  

By virtue of rotational invariance, the density of a tilted interface, as in
Fig.\ref{fig:sketch}.b will be still given by $\rhoi(\hper)$, but now $\hper$
will no longer be  a single function of the vertical distance $z-\ellav$, but will also depend on 
$\rpar$.
Whence,  in the absence of an external field, the only relevant direction in the 
hypothetical system of Fig.\ref{fig:sketch}.a-b 
is the perpendicular distance
to the interface, and  densities along that line are invariant to the choice of
reference frame.\cite{davis77,zia85}
%

In practice, the length scale relevant for the action of an external fields 
is often  much larger than the length scale of density correlations.  Such is the case 
of a liquid--vapor interface, where
the density profile decays in the length scale of a few angstrom, while the
capillary length, which sets the scale of action of gravity, is of the order of
the millimeter. Whence, the full density profile  may be described perturbatively
as that pertaining to a free interface, plus a small correction which will
depend on  the direction along the external field.

\subsection{Non-locality}

In practice, interfaces are not flat as in Fig.\ref{fig:sketch}.a, but rather, 
have a rough profile that results from thermal fluctuations
(Fig.\ref{fig:sketch}.c). 
Consider now a
hypothetical case were we could 
constrain
a given realization of the
film profile $\film(\rpar)$, with average $\prom{\film}{}=\ellav$.
Clearly, the resulting 
constraint density profile
$\rho(\rvec{}{};\funcL)$, can no longer be
expressed in terms of a single variable, but rather, depends on all three
Cartesian coordinates $\rvec{}{}$ (Fig.\ref{fig:sketch}.c). Likewise, $\rho(\rvec{}{};\funcL)$ can no
longer be expressed in terms of the simple intrinsic density, but rather,
must pick up a functional dependence of the full film profile, as indicated by the
second argument $\funcL$ of $\rho(\rvec{}{};\funcL)$. \cite{fisher94} Such must be the case when the film profile exhibits a
finite local curvature, $1/R$, for the observer locally will not be able to tell
whether that curvature corresponds to a fraction of a droplet or bubble 
of radius R, or
rather, to a piece of an undulated film profile \cite{mecke99b,blokhuis09}. Hence, the density in the
vicinity of the curved film must conform to the Laplace equation and deviate
from the  resulting planar interface. It is expected that such density
distortions could be described by the Laplacian of the film profile, at least
for small curvatures \cite{safran94,chaikin95}. 

If, however, the local radius of curvature is much larger  than the bulk
correlation length  and we are interested in the density at a point
$\rvec{0}{}=(\rpar_0,z_0)$ a distances much smaller than $R$ away from the 
interface, 
the fluid feels
locally a tilted film with no curvature (Fig.\ref{fig:sketch}.d).  
Following the 
arguments of the previous section,  the density along a line perpendicular to 
the film profile should then be approximately given in terms of 
$\rhoi$ and the single variable $\hper$, hence, as a local function of
$\hper$.\cite{davis77} 

Let $\rvec{\perp}{}=(\rpar_{\perp},\film(\rpar_{\perp}))$ be the point on 
$\film(\rpar)$
that is closest to $\rvec{0}{}$. The perpendicular distance of  $\rvec{0}{}$ to
the film profile may then be given as $\hper^2=\Delta\rpar^2
(1+m_{\perp})$, where $|\Delta\rpar|$ is the distance
between points $\rpar_{\perp}$ and $\rpar_0$ on the plane perpendicular to the 
$z$ axis; while $m_{\perp}$ is the slope of a vector
perpendicular to $\film$ at $\rvec{\perp}{}$. 

Clearly, the slope  $m_{\perp}$ is a local property of $\film$ at 
point $\rvec{\perp}{}$. If, however, one can describe 
$\film(\rpar)$ at $\rpar_{\perp}$ in terms of a Taylor expansion about 
$\rpar_0$ with sufficient accuracy, then we can give $\hper$ fully
as a local function of $\film$, $\nabla_{\rpar}\film$,
$\nabla_{\rpar}^2\film$, etc. at $\rpar_0$.  In this favorable case, we
can then describe the density profile as $\rhoi(\hper)$, hence, also as an
extended local function at $\rpar_0$ (Fig.\ref{fig:sketch}.d).

In the most general case, however, $\film(\rpar)$ at an arbitrary point can 
not be given as a Taylor expansion at sufficient distances away from $\rpar_0$.
Whence, the location of $\rpar_{\perp}$ and accordingly, the norm
$|\Delta\rpar|$ will become a highly nonlocal property, which can only be
determined if the full film profile is known all the way from $\rpar_0$ to
$\rpar_{\perp}$.\cite{mecke99b} Furthermore, there
could emerge several perpendicular distances to a given point, only one
corresponding to the shortest distance to that point. As a result, even 
if the density profile $\rho(\rvec{}{};\film)$ could be given in terms of
the intrinsic density profile, the function $\rhoi(\hper)$ would become a 
highly nonlocal function. 

The relevance of nonlocal effects on the density profile of rough interfaces 
has been emphasized at length by Parry and collaborators
\cite{parry06,parry07,bernardino09,parry12}. Such effects
are particularly important in the study of short range critical wetting, where
the external field is zero at all distances beyond the bulk correlation
length. 

In what follows, we will argue that for films subject to external fields of 
range greater than the bulk correlation length, an extended local 
approximation to the 
density profile is sufficient to capture the leading order corrections to the
classical capillary wave theory.

The origin of the extended locality is introduced in the theory under the
assumption that the density is a single variable of $\hper$. 
The observation that the density profile is best expressed as a
function of the perpendicular distance to the interface has already been
stressed previously \cite{davis77,diehl80,kawasaki82,zia85,stecki98,mecke99b,stecki01,blokhuis09}. 
In the next section we will show
that for small deviations away from planarity, $\hper$ may be expressed easily
in terms of a film profile and its gradient, and explore the consequences of
this assumption.

\section{Density profile}

In the classical theory of capillary fluctuations, the density of a 
rough instantaneous configuration at a point $\rvec{}{}$ is dictated merely 
by the vertical distance of that point from the film profile
$\film(\rpar{})$. Accordingly, the density profile $\rho(\rvec{}{};\funcL)$ 
may be
expressed in terms of an assumed {\rm intrinsic density profile}, $\rhoi(z)$ of
the single variable $\hver(z,\rpar{})=z-\film(\rpar{})$, as:
\begin{equation}\label{eq:classicalrho}
   \rho(\rvec{}{};\funcL) = \rhoi(\hver(z,\rpar{}))
\end{equation} 
In the previous section, however, we argued that in the low curvature limit, 
the density at a point should depend on the perpendicular distance to the
interface. Accordingly, the starting point of our study is to consider that we
can describe the full density still in terms of a function of $\rhoi(z)$, but,
with a more complex dependence given by the single variable $\hper(z,\rpar{})$. Whence, we will
henceforth explore the implications of the following ansatz:
\begin{equation}\label{eq:ansatz}
  \rho(\rvec{}{};\funcL) = \rhoi(\hper(z,\rpar{}))
\end{equation} 
As discussed above, this assumption will be accurate in the low curvature 
limit, where
$\hper(z,\rpar{})$ is then a purely local function of $\film$ and
$(\nabla_{\rpar}\film)^2$, 
\begin{equation}\label{eq:hpersmall}
  \hper(z,\rpar{}) = \frac{z-\film(\rpar)}{\sqrt{1 + (\nabla_{\rpar}\film)^2}}
\end{equation} 
\Eq{ansatz} together with \Eq{hpersmall} are the starting point of our theoretical approach.
Clearly, if we neglect the gradient, \Eq{classicalrho} and \Eq{ansatz} are equivalent
and the only significant fluctuations are given by the interface displacements
away from the average film profile $\delta\film(\rpar)=\film(\rpar)-\ellav$, as in
the classical theory \cite{buff65,weeks77}.

In fact, it has been shown that \Eq{ansatz} is a systematic low temperature
solution of the Landau-Ginzburg-Wilson Hamiltonian for a rough
interface.\cite{diehl80} Such
statement holds {\em exactly} to zeroth order, provided one defines the film 
profile $\film$ as a collective coordinate obeying the 
condition:\cite{diehl80,kawasaki82,huse85}
\begin{equation}
 \int \hat{\rho}(\rvec{}{}) \frac{d\rhoi(z-\film(\rpar))}{d z} dz = 0
\end{equation} 
As discussed recently, this definition of the film profile is closely
related to microscopic definitions employed to locate $\film(\rpar)$ in computer
simulation experiments, and are close to the optimal choice required to
extract the capillary wave signature from the spectrum of surface
fluctuations.\cite{hernandez-munoz16}

\subsection{Linearisation}

Although the ansatz embodied in \Eq{ansatz} allows us to remove the nonlocal
character of the constraint density profile, the problem is far more complex
than in the classical theory, since the variable $\hper$ can no longer
be interpreted as a translation of the interface position. As a result, an
expansion of $\hper$ about $z-\ell$ to quadratic order does not satisfy the
condition $\hper(z,\rpar)=0$ for $z=\film(\rpar)$. Ignoring this limitation,
it is still possible to expand $\hper$ to
quadratic order in the interface fluctuations, and express it
in terms of an effective translation about the average planar interface as
considered previously by Stecki:\cite{stecki98}
\begin{equation}\label{eq:htrans}
  \hper(z,\rpar{}) = \hverpi(z) - \delta \hper_{\pi}(z,\rpar{})
\end{equation} 
where:
\begin{equation}
\begin{array}{lll}
 \hverpi(z) & = & z - \ellav \\
 & & \\
 \delta \hper_{\pi}(z,\rpar{}) & = & \delta\film(\rpar{}) + 
      \frac{1}{2} \hverpi(z) (\nabla_{\rpar} \film)^2
\end{array}
\end{equation} 
Accordingly, we can assume that the full density profile is given in terms of
effective translations, exactly as in the classical capillary wave theory;
However, the
translation is here along a direction   perpendicular to the interface, rather 
than merely along the  vertical direction:
\begin{equation}
  \rho(\rvec{}{};\funcL) = \rhoi(\hverpi-\delta\hper_{\pi}(z,\rpar{}))
\end{equation} 
This result resembles a related approach by van Leeuwen and Sengers, who
hypothesized that the density profile could be given in terms of a compressed
shift of the interface position, rather than by a mere translation, i.e., they
assumed local displacements of the form $z-\alpha(z) \film$, with $\alpha(z)$
an undetermined compression factor which is evaluated a posteriori from
thermodynamic considerations.\cite{vanleeuwen89} A similar strategy
has been adopted by Robledo and Varea.\cite{robledo97} 
In our approach, the compression factor is
given directly in terms of the film profile gradient, and has a clear
physical origin.

Having written the normal distance in terms of a linearized normal translation,
we can now expand $\rho(\rvec{}{};\funcL)$ in powers of $\delta
\hper_{\pi}$ up to second order as:
\begin{equation}\label{eq:rhoexppi}
  \rho(\rvec{}{};\funcL) = \rhoi(z;\ellav) -
\frac{d\rhoi(z;\ellav)}{dz}\delta\hper_{\pi}(z,\rpar{}) + 
 \frac{1}{2}  \frac{d^2\rhoi(z;\ellav)}{dz^2}
  \delta\hper_{\pi}^2(z,\rpar{})
\end{equation} 
Since \Eq{htrans} is only accurate up to quadratic order in deviations about
$\funcL=\ellav$, we drop all higher order terms in the above result and are left
with the following equation:
\begin{equation}
  \rho(\rvec{}{};\funcL) = \rhoi(z;\ellav) -
\frac{d\rhoi(z;\ellav)}{dz}\delta\film(\rpar{}) - 
\frac{1}{2}(z-\ellav)\frac{d\rhoi(z;\ellav)}{dz}(\nabla_{\rpar}\film(\rpar{}))^2 + 
 \frac{1}{2}  \frac{d^2\rhoi(z;\ellav)}{dz^2}
  \delta\film^2(\rpar{})
\end{equation} 
The first, second and fourth terms of the right hand side are exactly as those
expected for the density profile of the classical capillary wave theory up to
second order. Extended capillary wave theories have emphasized the need to
account for terms in the Laplacian.\cite{mecke99b,blokhuis09} However,  our
study suggests the need to consider contributions on the film gradient. As
we shall see later, such terms feed into an effective surface tension at
a lower order than terms in the Laplacian.  The presence 
of next to leading order terms of order square gradient has long been
recognized,\cite{fisher94,stecki98} but its implications apparently not explored explicitly.

In practice, we will be considering external fields that are a function of $z$
only. In such case, the relevant property is the lateral average of the density
profile. Since linear terms in $\delta\film$ and $\nabla_{\rpar}^2\film$ 
vanish because of reasons of symmetry, we are then left with the following result:
\begin{equation}
  \prom{\rho(\rvec{}{};\funcL)}{\rpar} = \rhoi(z;\ellav) -
\frac{1}{2}(z-\ellav)\frac{d\rhoi(z;\ellav)}{dz}
\prom{(\nabla_{\rpar}\film)^2}{\rpar} +
 \frac{1}{2}  \frac{d^2\rhoi(z;\ellav)}{dz^2}
  \prom{\delta\film^2)}{\rpar}
\end{equation} 
In section \ref{cap:hamiltonians}, we will exploit this equation in order to estimate the free
energy cost of a rough interface subject to  an external field.  We  will show that the additional term 
in the square
gradient conveys information on the external
field to terms linear in the interface area. This will result in a coupling of
the effective surface tension to the external field.

For the time being, we notice that a thermal average of the density profile over
capillary wave realizations is formally equal to that performed laterally,
albeit with the lateral averages replaced by thermal averages:
\begin{equation}\label{eq:avrhot}
  \prom{\rho(\rvec{}{};\funcL)}{\Xi} = \rhoi(z;\ellav) -
\frac{1}{2}(z-\ellav)\frac{d\rhoi(z;\ellav)}{dz}
  \prom{(\nabla_{\rpar}\film)^2}{\Xi} +
 \frac{1}{2}  \frac{d^2\rhoi(z;\ellav)}{dz^2} 
  \prom{\delta\film^2}{\Xi}
\end{equation} 
This equation provides the capillary wave broadening density profile resulting
from \Eq{ansatz}. The first and third terms on the right hand side are
exactly as in the classical theory, but the second term provides
a capillary wave broadening contribution that depends
on the film gradient. This explicit dependence was identified
recently,\cite{macdowell13,macdowell14} but is implicit in an
older result by Davis.\cite{davis77}

\subsection{Modified `Convolution' Approximation}

At this point, it is interesting to note
that the small variable $\delta \hper_{\pi}$ has an average 
$\langle \delta \hper_{\pi} \rangle=\frac{1}{2}\hverpi \langle (\nabla_{\rpar}\film)^2\rangle$,
and to quadratic order in $\film$, has a variance
$\langle \delta\hper_{\pi}^2\rangle=\langle \film^2\rangle$. This suggests
that $\delta\hper_{\pi}$ could be considered a Gaussian random variable with
a non-zero average.

Taking this into account, one notices that
 \Eq{avrhot} may be considered as the result of
a "convolution approximation" with
a Gaussian Kernel
very much as in the classical theory.\cite{jasnow84} The difference
is that rather than considering a Gaussian distribution for vertical
displacements, $\hver$, we consider that it is the perpendicular displacements
$\hper$ which are Gaussian random variables, with a first moment
that is a function of the position $z$ along the interface. 
\begin{equation}\label{eq:convapp}
  \prom{\rho(\rvec{}{};\funcL)}{\Xi} = 
\frac{1}{
\sqrt{
2\pi \langle \film^2\rangle } }
  \int \rhoi(\hverpi - \delta\hper_{\pi}) \displaystyle \exp(-\frac{1}{2} 
 \frac{ (\delta \hper_{\pi} - \langle \delta \hper_{\pi} \rangle )^2 }
      { \langle \film^2 \rangle } ) \, d (\delta \hper_{\pi})
\end{equation} 
Clearly, by expanding $\rhoi(\hverpi - \delta\hper_{\pi})$ to second order
and performing the Gaussian averages, the above modified convolution recovers
\Eq{avrhot} exactly. 
Obviously, the truncation to second order is
only valid when the Gaussian Kernel is strongly peaked relative
to the interface width. This shows,
as expected, that the accuracy of \Eq{avrhot} is  limitted
to the case were  $\langle \film^2 \rangle$
is small compared to the bulk correlation length.

Notice that in principle it should be possible to calculate the distribution
of perpendicular distances by computer simulations and test whether it follows
Gaussian behavior.\cite{tarazona04}

\subsection{Scattering from a rough interface}

\label{sec:scattering}

The structure of a rough interface may be probed using
grazing angle x-ray or neutron  scattering.\cite{sinha88,braslau88}
For incident sources at
angles larger than the critical internal reflection, it suffices to
consider the first Born approximation, whence, we consider the intensity
of reflected radiation as:\cite{pershan12}
\begin{equation}\label{eq:born}
I({\bf Q}_{\rpar},Q_z) = \int d\rpar_1 d\rpar_2 d z_1 d z_2
 \left \langle \rho(\rpar_1,z_1) \rho(\rpar_2,z_2) \right \rangle
 e^{iQ_z (z_1 - z_2)} e^{i{\bf Q}_{\rpar}\cdot (\rpar_1 - \rpar_2)} 
\end{equation} 
Using the second order expansion for the density profile, \Eq{rhoexppi}, we can estimate
the density-density correlation function as:
\begin{equation}\label{eq:rhorho}
\begin{array}{lll}
  \left \langle \rho(\rpar_1,z_1) \rho(\rpar_2,z_2) \right \rangle & = &
 \rhoi(t_1)\rhoi(t_2) +
\frac{d\rhoi(t_1)}{d\ell}\frac{d\rhoi(t_2)}{d\ell} \left\langle \delta
\hper_{\pi}(\rpar_1)\delta \hper_{\pi}(\rpar_2)\right\rangle
+ \frac{1}{2} \rhoi(t_1) \frac{d^2\rhoi(t_2)}{d\ell^2}  \left\langle
\delta\hper_{\pi}^2(\rpar_1)\right\rangle \\ & & \\ & 
+ & \frac{1}{2} \rhoi(t_2) \frac{d^2\rhoi(t_1)}{d\ell^2}
\left\langle\delta\hper_{\pi}^2(\rpar_2)\right\rangle
- \rhoi(t_1) \frac{d\rhoi(t_2)}{d\ell}
  \left\langle\delta\hper_{\pi}(\rpar_2)\right\rangle 
- \rhoi(t_2) \frac{d\rhoi(t_1)}{d\ell}
  \left\langle\delta\hper_{\pi}(\rpar)\right\rangle
\end{array}
\end{equation} 
where we have employed $t_i=z_i-\ell$ for the sake of brevity.
By plugging this result for the correlation function into the Born
approximation, we find the spectrum splits into specular ($Q_{\rpar}=0$) and
diffuse ($Q_{\rpar}\neq 0$) contributions as (Appendix A):
\begin{equation}
  I({\bf Q}_{\rpar},Q_z) = I_{\rm spec}(Q_z) \delta({\bf Q}_{\rpar}) + I_{\rm diff}({\bf
Q}_{\rpar},Q_z)
\end{equation} 
The specular contribution provides information on height-height perpendicular
correlations of the interface, and is given by:
\begin{equation}\label{eq:specular}
\begin{array}{lll}
 I_{\rm spec}(Q_z) &  = &  \int d t_1 d t_2 \left [ 
 \rhoi(t_1)\rhoi(t_2)   + \rhoi(t_1)\displaystyle{\frac{d^2\rhoi(t_2)}{d\ell^2}}  \left\langle
\sum_{\bf q}\film^2({\bf q})\right\rangle \right . \\ & &  \\  & - & \left .
 \rhoi(t_1)\, t_2 \displaystyle{ \frac{d\rhoi(t_2)}{d\ell} } \left\langle
\sum_{\bf q} q^2 \film^2({\bf q})\right\rangle
\right ] e^{iQ_z (t_1 - t_2)}
\end{array}
\end{equation} 
The  diffuse contribution provides information of parallel correlations
of the film profile. It is given as:
\begin{equation}\label{eq:diffuse}
 I_{\rm diff}({\bf Q}_{\rpar},Q_z) = \int d t_1 d t_2
 \frac{d\rhoi(t_1)}{d\ell} \frac{d\rhoi(t_2)}{d\ell} \left\langle
 \film^2({\bf Q}_{\rpar})\right\rangle e^{iQ_z (t_1 - t_2)}
\end{equation} 
This results suggest that information on the film height fluctuations
may be extracted from the intensity of scattered radiation.
However, it is not possible to provide simplified expressions with out the
introduction of further approximations. In section \ref{sec:erf} we will introduce a
model which will allow us to obtain a more transparent interpretation of
specular and diffuse spectrum. 

At this stage it is convenient to remark two effects that have been neglected
and that obscure the 
interpretation of scattering experiments for large wave-vectors. 
1) Firstly,  the
splitting of purely perpendicular  and purely parallel correlations that occurs
in the specular and diffuse contributions to the scattering intensity is the
result of the linearisation of $\hper$, i.e., \Eq{htrans}. A coupling of 
terms in the film ($\film(\rpar)$) and film gradient ($\nabla \film(\rpar))$ 
occur both in the specular and diffuse 
contributions if we
retain the non-linearized form of $\hper$, \Eq{hpersmall}. 2) In the
approximations of \Eq{ansatz}, where the density is expressed as a function
of the intrinsic density profile, there  is implicitly a
pre-averaging of fluctuations with wavelength of the order of the bulk 
correlation length. Accordingly, the expressions above are only correct for
small wave-vectors, and will certainly break down for wavelengths of the order
of the bulk correlation length. For larger momentum transfer, the spectrum 
features a
coupling of transverse and longitudinal modes, as well as a coupling of
bulk-like and surface fluctuations, which make the interpretation of the results
very difficult and preclude the analysis of fine details of the capillary wave
fluctuations.\cite{gelfand90,paulus08,pershan12,parry16} 
3) A microscopic study of the density correlations of a fluid interface
for the Landau-Ginzburg-Wilson Hamiltonian
indicates that already for this simplified model the contributions from the interface feature not only
the leading order translation mode of the interface (which is correctly
identified with capillary-waves), but also additional surface contributions
which become important at large wave-vector transfer. Aside the bulk
correlations, the full spectrum may be expressed as a sum of Lorentzian
contributions.\cite{zittartz67} Whence, fitting the surface contributions by a single
Lorentzian  entangles the surface modes and obscures a clear interpretation
of the spectrum.

\subsection{Consistency checks}

\subsubsection{Consistency with renormalization group theory and scaling}

Let us now compare the  result of \Eq{avrhot} with expectations from 
 renormalization group theory in the one loop
approximation \cite{jasnow78,kopf08}. This approach has the advantage 
over capillary--wave theory that bulk and capillary wave fluctuations 
are treated ab--initio within a unified framework, so that the hypothesis of
an add-hoc intrinsic density profile is not implied a-priori. 
As a caveat, however, it should be noticed that the one-loop approximation
is unable to deal with strictly infrared divergences. Particularly, this
limitation holds for the well known translational goldstone mode
of the surface correlation function, which diverges as $1/q^2$, independently
of the distance away from the critical point. 
This limits severely
the scope of this theory, which becomes completely invalid for a free 
interface in the thermodynamic limit.
For practical  purposes, considering the interface under a pining field or within
a  finite system  provides a long wavelength cutoff that serves as a
mathematical
device to remedy the problem of infra-red divergences.\cite{jasnow84} 
Despite of this mathematical trick, the results  from the one loop approximation 
should be trusted only for surface fluctuations of the order of the bulk 
correlation length, wich effectively is the case when the pining field is strong
enough or the system size is small enough.

Baring this in mind, we consider results or the Landau-Ginzburg-Wilson
Hamiltonian, which exhibits the well known $\tanh(z)$ intrinsic density
profile. Jasnow and Rudnick first performed the calculation for a fluid
 under the gravitational field in the thermodynamic limit. K\"opf and 
M\"unster performed a related calculation for a fluid in a finite system of
lateral dimensions $L\times L$ and zero field. 
Whereas both results are found to be
consistent \cite{kopf08},  we choose here to show the result of  K\"opf and 
M\"unster, which is presented in a somewhat more readable form. 

Since
renormalization group calculations are usually performed in the language of the 
Ising model, we define a normalized density which ranges between $\pm 1$, as is
usual for the Ising magnetization:
\begin{equation}
   m(z) =
\frac{\prom{\rho(\rvec{}{};\funcL)}{\Xi}-\frac{1}{2}(\rho_l+\rho_v)}{\frac{1}{2}(\rho_l-\rho_v)}
\end{equation} 
where $\rho_v$ and $\rho_l$ are the vapor and liquid coexistence densities.
In terms of this normalized density, the thermally averaged density profile 
exhibits two distinct regimes. For large systems (or weak fields),
the interface roughening is large,  and the density magnetization is
given as a gaussian convolution of the intrinsic profile.\cite{jasnow84} 
For large roughness, the gaussian is very broad, the intrinsic features 
are washed out, and $m(z)$ becomes an error function, in agreement with
\Eq{convapp}.\cite{abraham81} Here, we are mainly interested in the
opposite limit of small systems or strong pining fields, where roughening is
small, and intrinsic features of the density profile remain recognizable
even close to the average interface position $z=0$. In that
case, the density profile is:\cite{kopf08}
\begin{equation}\label{eq:rgt}
\begin{array}{lll}
m(z) & = & 
 \tanh( \frac{1}{2}\frac{z}{\xi_R} ) + \displaystyle\frac{k_B T}{8\pi\gamma_R\xi_R^2}\left (
\alpha - \ln \frac{L}{\xi_R}  \right ) \tanh(\frac{1}{2} \frac{z}{\xi_R} )\,
\sech^2(  \frac{1}{2}\frac{z}{\xi_R} ) \\
& & \\
& & 
 - ( 3\ln3 - 13/4 ) \displaystyle\frac{k_B T}{32\pi\gamma_R\xi_R^3} \frac{z}{\xi_R} 
       \sech^2(  \frac{1}{2} \frac{z}{\xi_R} )
\end{array}
\end{equation}  
where a subindex "R" stands for the corresponding renormalized quantities,
and $\alpha=1.832$.

The first term in the right hand side corresponds to a mean field $tanh(z)$
density profile. This form follows because the one-loop approximation
has been worked out for the Landau-Ginzburg-Hamiltonian, with
the usual biquadratic free energy. A more complicated
form  could be obtained if one used an improved equation of state with built in 
critical exponents as in the Fisk-Widom theory.\cite{fisk69,rowlinson82b} 
Be as it may, it is found that the resulting  $tanh(z)$ intrinsic profile obeys 
the Fisk-Widom scaling hypothesis.


The second term does no longer conform to the scaling hypothesis, but rather,
 exhibits a logarithmic prefactor
which diverges very slowly, as $L\to\infty$. This divergence occurs
in the calculations because of the lack of a pining field. In the results by 
Jasnow, it is replaced by a logarithmic term in the gravitational field, which
exhibits an equivalent divergence in the limit where the field vanishes.
%
With or without external field,  the prefactor takes precisely the form
expected for interface position fluctuations as described by capillary wave
theory. Accordingly, it is  identified both in  Ref.\cite{jasnow78} and \cite{kopf08}
as a signature of capillary wave fluctuations, which appear naturally in this
theoretical framework with no a priori assumptions.  As a bonus of renormalization
group calculations, the spurious ultraviolet divergence of the interface fluctuations
which appears in  capillary wave theory is not an issue any longer.

The third term also does not conform to the scaling hypothesis, but has no clear
physical interpretation in the framework of renormalization group theory. 

However, motivated by \Eq{avrhot}, we realize that \Eq{rgt} may be actually written as:
\begin{equation}\label{eq:rgt-tuneada}
m(z) = 
 \tanh( \frac{z}{2\xi_R} ) + \frac{k_B T}{4\pi\gamma_R}(\ln\frac{L}{\xi_R} -
\alpha ) \frac{d^2}{dz^2} \tanh( \frac{z}{2\xi_R} )
 - (3\ln3 - 13/4)\frac{k_BT}{16\pi\gamma_R\xi_R^2} z \frac{d}{dz} 
       \tanh( \frac{z}{2\xi_R} )
\end{equation}  
whence,  the renormalization group results conform exactly to
\Eq{avrhot}, provided we assume a Fisk--Widom intrinsic density profile, and identify
the prefactors of $\tanh''$ and $z\,\tanh'$ with the mean squared fluctuations of
$\film$ and $\nabla_{\rpar}\film$, respectively. 

Since \Eq{rgt} is consistent with \Eq{avrhot}, and the latter is a systematic expansion of
the ansatz \Eq{ansatz}, it follows that the renormalization group result is actually
compatible with the following scaling form for the constrained magnetization:
\begin{equation}\label{eq:fiskwidom}
  m(\rvec{}{};\funcL) = \phi\left( \frac{\hper}{\xi_R} \right)
\end{equation} 
where $\phi$ is a suitable step like single variable function, while $m(\rvec{}{};\funcL)$ stands
here for a thermal average over bulk fluctuations consistent with the imposed
capillary wave constraint, $\funcL$. i.e., the Fisk--Widom scaling 
survives bulk--like fluctuations and holds at least for the constrained density
profile, provided the density is expressed in terms of the normal rather than
the vertical distance  to the interface.
The scaling form is lost only after thermally averaging over capillary waves,
but the significance of a collective coordinate for the intrinsic surface would
seem to hold up to the critical point, at least to the  accuracy of the
one loop approximation. Such separation of surface and bulk fluctuations is
consistent with the column model of the interface suggested by
Weeks,\cite{weeks77,huse85}
or the field theoretical calculations by Delfino and Viti.\cite{delfino12}
It also resembles previous work by van Leeuwen and Sengers, who  stressed
the need to introduce compressed shifts instead of mere displacements in order
to incorporate capillary wave fluctuations into the Fisk-Widom
theory.\cite{vanleeuwen89}

\subsubsection{Consistency with the Capillary Wave Hamiltonian}

Since the ansatz of \Eq{ansatz} was motivated from rather general symmetry
considerations, it is expected to hold irrespective of the particular
choice for $\rhoi$, or alternatively, of the assumed microscopic functional.

For convenience, let us consider here a free liquid--vapor interface, as
described by the square--gradient theory:
\begin{equation}\label{eq:sgt}
 A[\rho] = \int d\rvec{}{}\left \{ f(\rho) + \frac{1}{2} C \left (\nabla \rho 
\right )^2
\right \}
\end{equation} 
where $f(\rho)$ is some suitable local free energy. 

For a frozen realization of the film profile, we assume that the density is
given as the Euler-Lagrange equation:
\begin{equation}\label{eq:sgextr}
\frac{\partial f}{\partial \rho} - C \nabla^2 \rho(\rvec{}{};\film)  = 0
\end{equation} 

Assuming the ansatz \Eq{ansatz} for the extremal density, 
the second term of the above equation is readily written as:
\begin{equation}
  \nabla^2\rho(\rvec{}{};\film) =
 \frac{d^2 \rhoi(\hper)}{d \hper^2} (\nabla \hper)^2 +
 \frac{d \rhoi(\hper)}{d \hper} \nabla^2 \hper
\end{equation} 
with
\begin{equation}
\nabla \hper(\rvec{}{};\funcL) = \left \{
  \frac{\bf k}{\sqrt{1+(\nabla_{\rpar}\film)^2}} + 
\left ( 1 + \frac{(z-\film)\nabla_{\rpar}^2\film}{\sqrt{1+(\nabla_{\rpar}\film)^2}}  \right )
\frac{\nabla_{\rpar}\film}{\sqrt{1+(\nabla_{\rpar}\film)^2}}
\right \}
\end{equation}
Using this expression, and neglecting  higher order contributions in the
gradient and Laplacian, we find that $(\nabla \hper)^2$ is equal to unity. 
Accordingly, in the limit of small curvature that we are concerned, 
the extremal, \Eq{sgextr}, simplifies to:
\begin{equation}
\frac{\partial f}{\partial \rho} - C \frac{d^2 \rhoi(\hper)}{d \hper^2}  = 0
\end{equation} 
This equation may be integrated along the single variable $\hper$,
as in the standard Cahn-Hilliard theory of interfaces. 
We can then substitute the result into \Eq{sgt} and 
obtain a free energy which has an explicit functional dependence on 
$\funcL$:  
\begin{equation}
  A[\funcL] = \int d\rvec{}{} C \left (\frac{d\rhoi(\hper)}{d\hper}  \right)^2
\end{equation} 
where the dependence of $\hper$ on $\funcL$ has been omitted for the
sake of brevity. Considering that, for a free interface,
 the dependence of $\rhoi(\hper)$ on $\hper$ is
exactly as that on $z$, and performing a change of variables, 
the above result readily transforms into:
\begin{equation}
  A[\funcL]  = \int dz C \left (\frac{d\rhoi(z)}{dz}  \right)^2
              \int d\rpar \sqrt{1 + (\nabla_{\rpar}\film)^2}
\end{equation} 

Since the first integral may be immediately identified with the  mean field
surface tension, we find that the free energy now transforms exactly 
into the Capillary Wave Hamiltonian,
\begin{equation}\label{eq:cwhmf}
  \mathcal{H}[\funcL] = \gamma_{\rm cw} \int d\rpar \sqrt{1 + (\nabla_{\rpar}\film)^2}
\end{equation} 
with a bare surface tension, $\gamma_{cw}$ equal to the mean field surface 
tension of the van der Waals theory, i.e., \Eq{ansatz} is the approximate
expression for the density profile implied in the capillary wave Hamiltonian of
a free interface. This result was already anticipated by Davis under the
assumption that the extremal condition, \Eq{sgextr} is obeyed along the perpendicular
direction to the interface.\cite{davis77}

Using the method of collective
coordinates, Diehl et al. have shown the above result is as a systematic approximation 
to the renormalized solution of \Eq{sgt} which becomes exact in the low temperature limit (corresponding to
infinitely sharp interface with infinite surface tension).\cite{diehl80}

\section{Interface Hamiltonian}

Previously, we have discussed free interfaces. We now consider
how to extend the ansatz of \Eq{ansatz} to the special case
of wetting films adsorbed on a completely flat and structureless
 substrate that
is perpendicular to the $z$ direction. 
In such case, the interaction of the substrate with the fluid
may be described by means of an external field $V(z)$ which
only depends on $z$. Furthermore,
we will assume that the wetting film is sufficiently thick
that a liquid-vapor interface can still be identified as discussed in
section \ref{preliminary}. Accordingly, a film height for each
point $\rpar$ on the substrate may be defined as the distance
between the film profile $\film(\rpar)$ and the substrate.

Before continuing, let us mention that in the classical capillary
wave theory, the free energy of an adsorbed wetting film with
a corrugated liquid-vapor film profile $\film(\rpar)$ is
given by,\cite{rowlinson82b,evans92}
\begin{equation}\label{eq:ccwt}
   H[\film] = \int d\rpar \left \{ g(\film) + 
           \gamma \left( \nabla_{\rpar} \film \right )^2 \right \}
\end{equation} 
where, in our convention, $g(\film)$ is an unshifted interface potential, 
which bares all of the
free energy of the system for a completely flat adsorbed film.
Accordingly, in the limit of an infinitely thick film 
it becomes $g(\film\to\infty)=\gamma_{sl} + \gamma_{lv}$,
with $\gamma_{sl}$, the solid-liquid surface free energy, and
$\gamma_{lv}$,  the liquid-vapor surface tension.
The second contribution of the integral accounts for the cost of
increasing the liquid-vapor interfacial area. The coefficient
of the square gradient, $\gamma$, is an effective liquid-vapor surface tension
(also known as the stiffness coefficient in specialized literature).
In the classical capillary wave theory, $\gamma = \gamma_{lv}$. 

In this section, we use microscopic free energy functionals
in order to assess to what extent this equality is correct.


\label{cap:hamiltonians}
\subsection{Short range forces and external field}
\label{cap:srfef}

Let us now consider the case of an adsorbed liquid film, exhibiting
short range forces only. Particularly, let us assume that the
interactions of the fluid with the adsorbing substrate may be
described by a short range external field, $V_0(z)$, where the
subscript $0$ indicates here the short range nature of the field  
(and also anticipates  this system will be employed as a reference state
in a perturbation approach later on).

In the square gradient approximation, the free energy 
functional now reads:
\begin{equation}\label{eq:sgtv}
 A_0[\rho] = \int d\rvec{}{}\left \{ f(\rho) + \frac{1}{2} C \left (\nabla \rho 
\right )^2 + V_0(z)\rho
\right \}
\end{equation} 
In principle, the density profile of a rough
interface, with roughness $\funcL$, say  $\rho_0(\rvec{}{};\funcL)$,
 is obtained as the extremal of the
free energy functional, subject to the constraint  given by the film profile
$\film$. 
%
The stationarity condition amounts to the  usual 
partial differential equation:
\begin{equation}\label{eq:sgtext}
\frac{\partial f}{\partial \rho} - C \nabla^2 \rho_0 + V_0(z) = 0
\end{equation} 
together with an additional variational condition
at $z=0$ that fixes the density at the wall.\cite{jin93}

Unfortunately, solving this partial differential equation subject
to boundary conditions  is very difficult. At most, it is possible to 
find solutions for the mean field profile with flat liquid-vapor
interface,\cite{brezin83} 
which is identified with the intrinsic density profile of an adsorbed film
$\rhoi(z;\ell)$ of height $\ell$. 
%
In order to impose the variational condition at the wall,
the solution of \Eq{sgtext} must satisfy the full stationarity 
principle of \Eq{sgtv} in integral form, namely:
\begin{equation}\label{eq:sgtextint}
  \int  d \rvec{}{} \left \{ 
 \frac{\partial f}{\partial \rho}\,\delta \rho +
C \nabla \rhoi \cdot \nabla \delta \rho + V_0 \, \delta \rho 
\right \} = 0
\end{equation} 
where $\delta\rho$ is an arbitrary density variation (Appendix C).

Compared to the free interface, 
the presence of an external field very much complicates the solution
of \Eq{sgtext}, even for the mean field case,
since we can no longer assume that the intrinsic density profile is a 
function of $z-\ell$ alone. The sharp transition from liquid to vapor 
density will still be governed roughly by $z-\ell$, but the decay of wall
fluid--correlations must obviously depend essentially on the distance away from
the wall, which, assumed at the origin now yields an explicit dependence on
 $z$. For this reason, we must slightly generalize our ansatz, \Eq{ansatz} 
to deal with this complication. 

Considering that generally, the intrinsic density profile
of an adsorbed film is a function of $z$ and the interface position, $\ell$, we
now write:
\begin{equation}\label{eq:ansatzv}
  \rho_0(\rvec{}{};\funcL)=\rho_{\pi}(z;\ell=\film+\delta \hper)
\end{equation} 
where $\delta\hper$ is defined as the difference between the normal and
vertical distances to the interface, $\delta\hper=\hper-\hver$. In practice,
to the order of squared gradient terms in the film profile it amounts to:
\begin{equation}
  \delta\hper = \frac{1}{2} (z - \film) ( \nabla_{\rpar} \film )^2
\end{equation} 
Notice that the above result is fully equivalent to \Eq{ansatz} for the case 
where the intrinsic density profile only depends on the vertical distance 
$z-\ell$ and reduces to the Fisher-Jin ansatz in the limit where 
$\delta\hper\to 0$.\cite{fisher94} Physically, it assumes that the relevant
film height required to describe the density at a point is given
as the distance to the substrate along the normal to the interface.
This obviously cannot possibly be exact, and will fail close to the substrate.
However, it is very accurate close to the liquid-vapor
interface.\cite{nold16} Since,
in practice, large density gradients occur mainly  at this interface, the approximation
is justified.

In order to calculate the free energy, we now substitute the above result
into the square gradient functional, whence:
\begin{equation} \label{eq:sgtansatz}
 A_0[\rho_0(\rvec{}{},\funcL)] = \int d\rvec{}{}\left \{ f\left(
 \rhoi(z;\film+\delta\hper)\right) + \frac{1}{2} C \left (\nabla 
 \rhoi(z;\film+\delta\hper) \right )^2 + V_0(z)\rhoi(z;\film+\delta\hper)
\right \}
\end{equation} 
Despite the simplifying assumption embodied in \Eq{ansatzv}, we  find that 
transforming the 
Cahn-Hillard functional into an Interface Hamiltonian can only be performed
exactly in the limit of small gradients
$\sqrt{1+(\nabla_x\film)^2}\to 1+\frac{1}{2}(\nabla_x\film)^2$, and even so only to
order $(\nabla_x\film)^2$. The reason is that making the change of variables
that was convenient in the absence of an external field  makes the external
field a function of $\film$ and its gradient, so one cannot get rid of
this complicated dependence by changing variables.

For this reason, we can only proceed by performing an expansion of the density
profile in powers of $\delta\hper$, to first order:
\begin{equation}\label{eq:rhoexpds}
 \rhoi(z;\film+\delta\hper) = \rhoi(z;\film) + \frac{\partial
\rhoi(z;\film)}{\partial \ell}\,
\delta \hper
\end{equation} 
Substitution of this result into the first two contributions of \Eq{sgtansatz}, 
followed by a Taylor expansion, we find (Appendix B):
\begin{equation}\label{eq:fexpds}
f\left(\rhoi(z;\film+\delta\hper)\right) =
 f(\rhoi(z;\film)) + 
 \frac{\partial f}{\partial \rho} \frac{\partial \rhoi(z;\film)}{\partial \ell} \delta \hper
\end{equation} 
\begin{equation}\label{eq:gradexpds}
\begin{array}{lcl}
\left ( \nabla \rhoi(z;\film+\delta\hper) \right )^2 & = &
\displaystyle{
\left (\frac{\partial \rhoi(z;\film)}{\partial z} \right )^2 +
2 \frac{\partial \rhoi(z;\film)}{\partial z} \frac{\partial^2
\rhoi(z;\film)}{\partial z \, \partial \ell}
 \delta \hper } \\ & & \\
 & + & 
\displaystyle{ \left [ \left (\frac{\partial \rhoi(z;\film)}{\partial \ell} \right )^2 +
 \frac{\partial \rhoi(z;\film)}{\partial z} \frac{\partial
\rhoi(z;\film)}{\partial \ell} \right ] 
 \left ( \nabla_{\rpar} \film \right )^2 }
\end{array}
\end{equation} 
By replacing \Eq{rhoexpds}--\Eq{gradexpds}  into \Eq{sgtansatz}, 
and collecting terms of order
$(\nabla_{\rpar}\film)^2$,  the free energy can be
expressed as  a linearized interface Hamiltonian: 
\begin{equation}\label{eq:hamsrsg}
  H_0[\funcL] = \int d \rpar{}{} \left \{ g_0(\film) + 
 \frac{1}{2} \gamma_0(\film) \left ( \nabla_{\rpar} \film \right )^2 \right \}
\end{equation} 
The local free energy, $g(\film)$, contains terms that are independent of
the film gradient and may be readily identified with the interface potential 
of a flat film of height $\film$:
\begin{equation}
 g_0(\film) = \int d z \left \{ f(\rhoi(z;\film)) 
 + \frac{1}{2} C \left (\frac{\partial \rhoi(z;\film)}{\partial z} \right )^2
 + V_0(z) \rhoi(z;\film)  \right \}
\end{equation} 
Notice that in our definition, $g_0(\ell\to\infty)=\gamma_{sl}+\gamma_{lv}$.
The effective surface tension, $\gamma_0(\film)$, with an explicit film height
dependence, contains those terms from \Eq{rhoexpds}-\Eq{gradexpds} which are 
factors of the film gradient: \begin{equation}\label{eq:gammasgtv0} \begin{array}{lcl} \gamma_0(\film) & = & \displaystyle \int d z \left \{ 
 \left [  \frac{\partial f}{\partial \rho} + V_0(z)  \right ] 
  ( z - \film ) \frac{\partial \rhoi(z;\film)}{\partial \ell}  
\right . 
\\ & & \\
 & + & 
\displaystyle 
\left .  
 C \frac{\partial \rhoi(z;\film)}{\partial z} \frac{\partial\,}{\partial z}
\left [  
        (z-\film) \frac{\partial \rhoi(z;\film)}{\partial \ell}
\right ]
+
C  \left (\frac{\partial \rhoi(z;\film)}{\partial \ell} \right)^2 
\right \} 
\end{array}
\end{equation} 
%
In order to simplify the above result for  $\gamma_0(\film)$, 
we notice that the first three terms of the right hand side
obey the stationarity condition of
the intrinsic density profile, \Eq{sgtextint}, for the particular
choice $\delta \rho=(z-\film)\frac{\partial\rhoi}{\partial\ell}$. 
Since \Eq{sgtextint} holds for arbitrary density variations,
it follows that these three terms  cancel each
other exactly, and only the last term of \Eq{gammasgtv0} survives:
\begin{equation}\label{eq:gammasgtv1}
 \gamma_0(\film) = \int C 
  \left (\frac{\partial \rhoi(z;\film)}{\partial \ell} \right)^2 d z 
\end{equation} 
The above result corresponds to the position 
dependent stiffness of the Fisher-Jin theory.\cite{jin93} 
It provides corrections to
the surface tension that arise mainly from the distortion of
the liquid-vapor interface by the  substrate. 
Accordingly, for short range systems in a Cahn-Hillard approximation,
\Eq{ansatzv}, provides exactly the Fisher-Jin Hamiltonian, wich merely is
the result for the approximation of \Eq{ansatzv} with neglect of $\delta s$.
It follows that our ansatz provides exactly the same predictions
for short range wetting as the Fisher-Jin theory. Particularly,
it suffers from a stiffness instability close to the critical wetting
transition that seems inconsistent with simulations.\cite{bryk13}
Therefore, it does not seem that this approach can shed any new light on
this difficult problem. In such cases,  as will be discussed shortly
for systems in a long range field,
the more elaborated Non-Local model should be preferred.\cite{parry06,parry07}

\subsection{Short range forces and a long range external field}

\label{cap:srlr}

Although not stated explicitly, the above results are 
in principle
only valid for
fluids under short range external fields. Indeed, the ansatz of \Eq{ansatzv},
implying a dependence of density on the perpendicular distance to the film holds
strictly in an isotropic  system, as discussed in section \ref{preliminary}. 
Furthermore, use of \Eq{gammasgtv1} requires knowledge of the exact
intrinsic density profile of a fluid under an external field, which
is available usually only 
for external fields of very short range.\cite{brezin83,jin93} 


The above results are still useful, because we can exploit them as a
reference system in a perturbation approach. Whence, consider again a
fluid with short range forces, which, subject to the short range
external field $V_0(z)$ is well described by the free energy functional
 of \Eq{sgtv}. Let us now assume that, on top of the external field we
allow for a long range perturbation, $V(z)$. The full free energy functional
is then well described as:
\begin{equation}\label{eq:funclr}
 A[\rho] = A_0[\rho] + \int V(z) \rho(\rvec{}{}) d \rvec{}{}
\end{equation}  
where $A_0$ stands for the free energy functional of \Eq{sgtv}. Let us now
assume that the density profile of the full Hamiltonian, $\rho$, may be
described without loss of generality
 as $\rho=\rho_0 +  \delta\rho$, where $\rho_0$ is
the density profile which extremalizes $A_0$. Then, plugging this series
into \Eq{funclr}, and expanding about $\rho_0$, yields, to first order:
\begin{equation}\label{eq:perturbativamente}
 A[\rho] = A_0[\rho_0] + \int  V(z) \rho(\rvec{}{}) d \rvec{}{} + O[(\delta \rho)^2]
\end{equation} 
As noted by Parry and coworkers,\cite{parry07,bernardino09}
the reference free energy functional does not contribute to the free energy
at first order in the perturbation, because $\rho_0$ is an extremal of $A_0$.

This result is still not convenient, because  it is given in terms of the
unknown density, $\rho$. However, for adsorbed liquid films 
the perturbation due to an external field is of order $\delta \rho \propto 
\rho_l \kappa_l V(z)$, where $\rho_l$ and $\kappa_l$ are the bulk liquid
density and bulk liquid compressibility, 
respectively.\cite{barker82,dietrich91,macdowell14} Whence, for liquids 
below the critical point, which are highly
incompressible, the perturbation is very small, and the zeroth order
approximation $\rho \approx \rho_0$ is very good.

Accordingly, we merely need to replace \Eq{ansatzv}, 
into \Eq{perturbativamente}. The free energy in excess to
the reference state is given by:
\begin{equation}\label{eq:bindingp}
W[\film] = \int V(z) \rhoi(z;\film+\delta\hper) d\rvec{}{}
\end{equation}  
Unfortunately, the resulting expression does not follow exactly the 
usual form of an Interface Hamiltonian (i.e., it does not split into a local 
interface potential and a surface tension term). This problem
has been emphasized by Parry et al.\cite{parry06,parry07,bernardino09} 
They note than an Interface Hamiltonian must rather be
described in terms of a Binding Potential which is of Non-Local
nature (i.e., it cannot be given merely as a local function of $\film$). 
The attempt to linearize this potential into the form of a 
classical Interface Hamiltonian fails to describe correctly the
wetting properties of strongly fluctuating systems.\cite{parry06,parry07}

In what follows, we shall be concerned only with fluids subject to
strong adsorption. Thus, the fluctuations are severely reduced by
the external field, and the Binding Potential $W[\film]$ may be linearized
savely. This can be achieved by replacing \Eq{rhoexpds} into
\Eq{bindingp} and \Eq{perturbativamente}, with the result:
\begin{equation}
  H[\funcL] = \int d \rpar{}{} \left \{ g(\film) + 
 \frac{1}{2} (\gamma_0(\film) + \Delta\gamma(\film)) 
          \left ( \nabla_{\rpar} \film \right )^2 \right \}
\end{equation} 
where we have identified:
\begin{equation}
 g(\film) = g_0(\film) + \int V(z) \rhoi(z;\film) d z
\end{equation} 
and,\cite{benet14b}
\begin{equation}\label{eq:gammafilmv}
 \Delta\gamma(\film) = \int V(z) (z-\film) \frac{\partial
\rhoi(z;\film)}{\partial \ell}  dz
\end{equation} 
Thus, apart from the short range dependence of the surface tension, $\gamma_0$,
systems with a long range external field will exhibit also 
an explicit dependence on $V(z)$ that was overlooked by Davis.\cite{davis77}
%
However, it must be born in mind that the  this effective
surface tension has its origin in the Non-Local Binding Potential. i.e.,
it is more akin to the external field than it is to the liquid-vapor
interface.

The explicit result of \Eq{gammafilmv} relies on the linearization
of the density profile (c.f. \Eq{rhoexpds}), and this requires a word
of caution.\cite{parry16b} From the form of \Eq{bindingp}  it is clear that
the factor of $V(z)$ inside \Eq{gammafilmv} should  decay as $\rhoi(z)$.
However, the long range decay that results after linearization is 
rather $z\,\partial \rhoi(z)/\partial\ell$. For systems under
long range external forces, which is our main concern here,
 $\rhoi(z)$ decays algebraically as $z^{-3}$,\cite{barker82} and 
the linearization
does not upset the correct asymptotic decay.
For systems with only short range forces, the leading order decay 
for $\rhoi(z)$ is exponential, whence, the linearization does not preserve
the correct long tail behavior.\cite{parry16b} In such cases, it may be 
required
to retain the form of $W[\film]$ without linearization. 
However or checks with an exactly solvable model indicate
that the approximation remains correct up to linear order in
the external field even for density profiles with an exponential decay.
Such checks also show that if the exact density profile for the system
in the external field is available, then the perturbative result of
\Eq{gammafilmv} is consistent with \Eq{gammasgtv1}.
(c.f. section \ref{sec:exact}).
 At any rate,
our phenomenological approach is most likely unreliable for strongly
fluctuating interfaces, and the full Non-Local theory should be
prefered in that case.\cite{parry06,parry07}

Finally note that the dependence of the surface tension on film height, both as
given in \Eq{gammasgtv1} and \Eq{gammafilmv}, is explicitly dependent on the choice of dividing
surface, since there is an explicit dependence in $\film$.\cite{parry06} 
This is not
altogether surprising, since  the surface area of a curved interface
depends on  an arbitrary choice  of the interface position, as largely
discussed in studies of nucleation and surface thermodynamics.\cite{ono60}
Previously, Blokhuis has stressed the dependence of the bending rigidity
coefficient on the choice of interface position.\cite{blokhuis09}

\subsection{Long range forces and an adsorbing wall}

Dealing with long range fluid-fluid forces is far more complicated. The reason
is that the gradient expansion that leads to the local Square Gradient 
functional does not converge in this case.\cite{evans92} Accordingly, it is 
necessary to resort to a van der
Waals functional that features explicitly the fluid--fluid pair potential,
$u(\rvec{21}{})$, with $\rvec{21}{}=\rvec{2}{}-\rvec{1}{}$:
\begin{equation}\label{eq:helvdw}
  A_{\rm vdw}[\rho] = 
 \int f(\rho) d\rvec{1}{} + 
\frac{1}{2} \int\int u(\rvec{21}{})
\rho(\rvec{1}{})\rho(\rvec{2}{}) d\rvec{1}{} d\rvec{2}{}
+ \int V_0(z) \rho(\rvec{1}{}) d \rvec{1}{}
\end{equation} 
The double integral over the pair interactions makes 
this functional  less amenable to analytical calculations, but, more
importantly, implies the need to introduce a wave-vector dependent surface
tension,\cite{napiorkowski93,mecke99b,blokhuis09,chacon14} as we shall see
shortly.

In principle, the optimal density profile $\rho(\rvec{}{};\funcL)$ must obey 
the extremal condition,  which, for this functional has the form of an integral
equation:
\begin{equation}\label{eq:extrvdw}
     \frac{\partial f(\rho)}{\partial \rho} 
  + \int u(\rvec{21}{})
 \rho(\rvec{2}{}) d\rvec{2}{} + V_0(\rvec{1}{}) = 0
\end{equation} 
Solving this equation analytically is already impossible for a flat film
$\film(\rpar)=\ellav$, whence, we cannot hope to obtain solutions for  the rough
interface. 

Again, we assume {\em a priori} that the extremal density obeys our
ansatz \Eq{ansatzv} for the density profile.
In order to avoid mathematical complications as much as possible, we 
expand the density profile to first order about $\film$, as in \Eq{rhoexpds}.
Quite generally, we can then write the free energy as a first order
density functional expansion:
\begin{equation}
  A[\rho(\rvec{}{};\funcL)] = A[\rhoi(z;\film)] + 
      \int \left . \frac{\delta A[\rho]}{\delta \rho(\rvec{}{})}
 \right |_{\rhoi(z;\film)} \delta\rho(\rvec{}{}) d \rvec{}{}
\end{equation} 
Notice that the  integrand of the second term in the right hand side does not
vanish, because $\rhoi(z;\film)$ is {\em not} a solution of \Eq{extrvdw}.
However, the first functional derivative does indeed vanish for the intrinsic
density profile of the flat film, $\rhoi(z;\ellav)$. It follows that the
integrand is at least of order $\film$, while, from \Eq{rhoexpds}, $\delta \rho$
is of order $\film^2$. Accordingly, the zero order solution:
\begin{equation}
    A[\rho(\rvec{}{};\funcL)] = A[\rhoi(z;\film)]
\end{equation} 
is exact to order $\film^3$. 

This rather general argument explains why our apparently complicated ansatz,
\Eq{ansatzv} reduces to the Fisher-Jin Hamiltonian for the case of short range
forces (c.f. Section \ref{cap:srfef}). The simplification at this stage allows us to
avoid very lengthy algebra in this case, and makes the problem tractable. Our
task is now merely to extend the approach of Napiorkowski and Dietrich to the 
case
of an adsorbed film.  Accordingly, we substitute $\rhoi(z;\film)$ in
\Eq{helvdw} to get:
\begin{equation}
\begin{array}{lcl}\label{eq:helvdwbabe}
 A_{\rm vdw}[\rho(\rvec{}{};\funcL)] & = & 
\displaystyle \int f( \rhoi(z;\film(\rpar{}) )) d\rvec{}{}
    + \frac{1}{2} \int\int  u(\rvec{21}{})
 \left  [   \rhoi(z_1;\ellav) \rhoi(z_2;\ellav) +  \right .  \\ & &
\\ & &
\displaystyle 2 \rhoi(z_1;\ellav) \frac{\partial \rhoi(z_2;\ellav)}{\partial \ellav}
\delta \film(\rpar_2) + \rhoi(z_1;\ellav) \frac{\partial^2
\rhoi(z_2;\ellav)}{\partial \ellav^2} \delta \film^2(\rpar_2) +
 \\ & & \\ & &
\left .
\displaystyle
 \frac{\partial \rhoi(z_1;\ellav)}{\partial \ellav}
 \frac{\partial \rhoi(z_2;\ellav)}{\partial \ellav} \delta \film(\rpar_1)
 \, \delta \film(\rpar_2)
 \right ] d\rvec{1}{} d\rvec{2}{} + \int V_0(z) \rhoi(z;\film(\rpar{})) d\rvec{}{}
\end{array}
\end{equation}
In order to arrange this expression into an interface potential (proportional
to the projected area) and a surface term (proportional to the interface
area), we write for the product of film heights:
\begin{equation}
\delta \film(\rpar_1) \, \delta \film(\rpar_2) = 
\frac{1}{2} \left \{  \delta \film(\rpar_1)^2 +  \delta \film(\rpar_2)^2 -
 ( \delta \film(\rpar_2)- \delta \film(\rpar_1) )^2 \right \}
\end{equation} 
Replacing this into \Eq{helvdwbabe}, we find that the free energy can be 
cast as:
\begin{equation}\label{eq:avdw3}
  A_{\rm vdw}[\funcL] = \int g(\film(\rpar)) d\rpar -
     \frac{1}{4} \int\int u(\rvec{21}{}) \frac{\partial \rhoi(z_1;\ellav)}{\partial \ellav}
 \frac{\partial \rhoi(z_2;\ellav)}{\partial \ellav}
      ( \film(\rpar_2)- \film(\rpar_1) )^2 d\rvec{1}{}d\rvec{2}{}
\end{equation} 
with the interface potential identified as:
\begin{equation}\label{eq:glr}
\begin{array}{lcl}
 g(\film) & =  &
\displaystyle
 \int f( \rhoi(z;\film )) d z
    + \frac{1}{2} \int\int  u(\rvec{21}{})
 \left [ \rhoi(z_1;\ellav) \rhoi(z_2;\ellav) +  
\displaystyle 2 \rhoi(z_1;\ellav) \frac{\partial \rhoi(z_2;\ellav)}{\partial \ellav}
\delta \film(\rpar) + 
\right .
\\ & &
\\ & &
\displaystyle
\left .
\left( \rhoi(z_1;\ellav) \frac{\partial^2
\rhoi(z_2;\ellav)}{\partial \ellav^2}  +
 \frac{\partial \rhoi(z_1;\ellav)}{\partial \ellav}
 \frac{\partial \rhoi(z_2;\ellav)}{\partial \ellav} \right ) \delta \film(\rpar)^2
  \right ] d z_1 d z_2 d\rpar_{21} + \int V_0(z)  \rhoi(z;\film) d z
\end{array}
\end{equation} 
Notice that the contributions explicit in the pair potential are approximated as
a second order expansion about $\film$. In practice, all terms linear in
$\delta\film$ vanish because of the extremal condition for the intrinsic density
profile.

The crucial difference between long and short range forces lies in the second
term of \Eq{avdw3}, which corresponds to the free energy cost for roughening the
interface. For the van der Waals functional, it is not explicitly a function 
of the film 
height gradient. The consequence is that it is not possible to decouple
the film height fluctuations from the pair potential. Of
course, powers of the gradient could appear explicitly by  expanding 
$\film(\rpar_2)$ about $\film(\rpar_1)$. Unfortunately, such expansion involves
moments of the pair potential which are {\em not} convergent for long range
forces.\cite{evans92} 

The way out is to manipulate the double integral of \Eq{avdw3}
in a similar fashion
as performed for the calculation of the structure factor (c.f. section
\ref{sec:scattering} and Appendix A), 
by replacing $\film(\rpar)$ with its Fourier representation. After some
additional calculations, it is possible to arrive at an expression
for the Interface Hamiltonian in Fourier space:
\begin{equation}\label{eq:hqvdw}
 H_{\rm vdw}[\funcL] = A g(\ellav) + \frac{1}{2} A \sum_{\qvec} \left [
              g''(\ellav) + \gamma_{\rm vdw}(\ellav;q) q^2 
 \right ] \film^2(q) 
\end{equation} 
where $\qvec$ is a wave-vector in the reciprocal space of $\rpar$,
 $g''$ is the $\ellav$ derivative of the interface potential,
\Eq{glr}. Because of the coupling of the pair potential with the film
fluctuations, the only way of writing a free energy that conforms to
the capillary wave theory, is by admitting an extra wave-vector dependence
into the surface tension:
\begin{equation}\label{eq:ihlrf}
 \gamma_{\rm vdw}(\ellav;q) = \int\int 
 \frac{\partial \rhoi(z_1;\ellav)}{\partial \ellav}
 \frac{\partial \rhoi(z_2;\ellav)}{\partial \ellav} 
  \left [ \frac{u(z_{21};q)-u(z_{21};q=0)}{q^2} \right ] d z_1 d z_2
\end{equation} 
where $u(z_{21};q)$ is the lateral Fourier transform of the pair potential.
This result is the generalization of a result due to
Blokhuis for free interfaces.\cite{blokhuis09} 

In systems with short range forces,  it is possible to 
make an expansion in even powers of $q$
and truncate to second order. To this order of approximation,
 $\gamma(\ellav;q)$  bares no explicit $q$ dependence, and becomes equal to 
the square gradient result for the surface tension, \Eq{gammasgtv1}. In this
case,  \Eq{hqvdw} merely becomes the Fourier representation for the
Interface Hamiltonian of a system with short range forces, \Eq{hamsrsg}.

The situation is different when we deal with long range forces, because
then $u(z_{21};q)$ may exhibit a weak logarithmic singularity. Particularly, 
for systems with dispersion forces:
\begin{equation}
      u(r) = -C_6/r^6
\end{equation} 
the lateral Fourier transform is, to leading
order:\cite{mecke99b,blokhuis09,chacon14}
\begin{equation}
     u(z;q)-u(z;q=0) = u_2(z)\, q^2 + 
  \frac{\pi C_6}{32}\, q^4 \ln(q R) + u_4(z)\, q^4 + O(q^6)
\end{equation} 
where $u_2(z)$ and $u_4(z)$ are the second and fourth derivatives of $u(z;q)$ 
with respect
to $q$, while $R$ is a constant of order the molecular diameter.

Using this expansion, one finds that the surface tension has the form:
\begin{equation}\label{eq:gammaqvdw}
  \gamma_{\rm vdw}(\ell;q) = \gamma_0(\ell) + \mu(\ell) q^2 \ln (q R) +
\kappa(\ell) q^2
\end{equation} 
with
\begin{equation}\label{eq:gamma0vdw}
 \gamma_0(\ell) =\int\int \frac{\partial \rhoi(z_1;\ellav)}{\partial \ellav}
 \frac{\partial \rhoi(z_2;\ellav)}{\partial \ellav} u_2(z_{21}) d z_1 d z_2
\end{equation} 
\begin{equation}
 \mu(\ell) =  \frac{\pi C_6}{32} \int\int 
 \frac{\partial \rhoi(z_1;\ellav)}{\partial \ellav}
 \frac{\partial \rhoi(z_2;\ellav)}{\partial \ellav}  d z_1 d z_2
\end{equation} 
and
\begin{equation}\label{eq:kappavdw}
 \kappa(\ell) =\int\int \frac{\partial \rhoi(z_1;\ellav)}{\partial \ellav}
 \frac{\partial \rhoi(z_2;\ellav)}{\partial \ellav} u_4(z_{21}) d z_1 d z_2
\end{equation} 
These equations are  again
a  generalization of the result expected for the
free interface of a fluid with van der Waals
forces.\cite{napiorkowski93,mecke99b,blokhuis09}
Alternatively, they may be considered  a generalization of
results of adsorbed interfaces with short range forces,\cite{parry94} to the
case of long range forces. Recall also that
the expression for the bending rigidity, $\kappa(\ellav)$
is incomplete, since we have ignored from the start curvature terms
which contribute terms of order $q^4$ into 
$\gamma(\ellav;q)$.\cite{mecke99b,blokhuis09}

\subsection{Long range forces and a long range external field}

Accounting for the effect of long range wall--fluid interactions
is now an easy problem, since we can proceed exactly as in section 
\ref{cap:srlr}, by
considering the Hamiltonian of \Eq{helvdw} as a reference system, and
the influence of the long range field as a perturbation. The resulting
Hamiltonian has the form of \Eq{hqvdw}, with a 
surface tension which
is the sum of  \Eq{gammaqvdw} and \Eq{gammafilmv}. 

\subsection{Summary}

\label{subsec:summary}

Before ending this lengthy section, it will be convenient to summarize the
results for later use. In essence, using the ansatz \Eq{ansatzv} for the density
profile of an adsorbed liquid film of height $\ellav$, we find that the
free energy of a rough realization of the film profile may be generally
given as:
\begin{equation}\label{eq:genfree}
 H[\funcL] = A g(\ellav) + \frac{1}{2} A \sum_{\qvec} \left [
              g''(\ellav) + \gamma(\ellav;q) q^2 
 \right ] \film^2(q) 
\end{equation} 
where $g(\ellav)$ is the interface potential, $g''(\ellav)$ is its second
derivative with respect to $\ellav$, and $\gamma(\ellav;q)$ is a wave-vector and
film height dependent surface tension. In the most general case it may be
written as:
\begin{equation}\label{eq:ggamma}
 \gamma(\ellav;q) =   \gamma(\ellav) +
 \mu(\ellav) q^2\ln(qR) + \kappa(\ellav) q^2 + O(q^4)
\end{equation} 
where $\gamma(\ellav)$ is the zero wave-vector surface tension:
\begin{equation}\label{eq:ggammaq0}
 \gamma(\ellav) =
 \gamma_0(\ellav) + \Delta\gamma(\ellav) 
\end{equation} 
The leading order coefficient, $\gamma_0(\ellav)$, may be interpreted
as a generalized surface tension that smoothly tends to the liquid-vapor surface
tension, $\gamma_{\rm lv}$, as film height increases. The origin of
the film height dependence is the distortion of the liquid-vapor density
profile in the neighborhood of the substrate.
It is given by \Eq{gammasgtv1} in the square gradient approximation,
or by \Eq{gamma0vdw} in the van der Waals approximation.  The next contribution,
$\Delta\gamma(\ellav)$ stems from the long range interaction of the
substrate on the liquid-vapor profile, and is given by \Eq{gammafilmv}, whether we conform
to the square gradient or the van der Waals approximation. The contribution
that is a factor of $\mu(\ell)$ is a singular term that results from
the presence of dispersive interactions, and vanishes altogether for
short-range forces. Finally, $\kappa(\ellav)$ is the bending rigidity, and
here it is given by \Eq{kappavdw}. It is finite whether the interactions are short or
long range, but vanishes within the square gradient approximation. 
Recall once more, however, that a more rigorous study shows that
density functional approaches based on phenomenological models for the
density profile are unable to provide the correct physics for effects
of order $q^2$ in the surface tension.\cite{parry16} Bearing this in mind,
we will nevertheless retain the term of order $q^4$ and consider 
$\kappa(\ellav)$ as a phenomenological coefficient. Notice that depending
on the choice for the surface location the sign of $\kappa(\ellav)$ may
be either positive or negative, but it has been shown that consistent
definitions for the surface location provide bending rigidities that
are positive.\cite{chacon05,tarazona07,hofling15}

The free energy in \Eq{genfree} is quadratic in the Fourier 
modes, equipartition
of energy holds exactly to this order of approximation, and the spectrum
of fluctuations follows immediately as:
\begin{equation}\label{eq:ncws}
 \prom{\film^2(q)}{\Xi} = 
\frac{k_B T}{
\left [
              g''(\ellav) + \gamma(\ellav;q) q^2 
 \right ] A }
\end{equation} 
This result is an improved expression for the spectrum of surface fluctuations
in the presence of an external field.\cite{rowlinson82b} Relative to the classical result,
the external field not only provides a low wave-vector bound
to the surface fluctuations, but also modifies the coefficient of $q^2$ by
an amount $\Delta\gamma$ which we will see, may be related to $g''(\ellav)$
for systems subject to a long range external field.

%
%

From  the results of section \ref{sec:scattering}, 
the surface spectrum
is accessible in principle via the study of density fluctuations
as determined from the structure factor.\cite{pershan12}
In practice, for reasons mentioned before it is difficult
to single out purely capillary-wave contributions in x-ray
scattering experiments. Rather, computer simulations
seem a more adequate means of testing fine feature of
the surface structure.\cite{chacon03,chacon05,tarazona07}
Indeed, recent computer simulations of the spectrum of
surface fluctuations provide strong evidence in support
of \Eq{ncws}.\cite{macdowell13,macdowell14,benet14b,chacon14,hofling15}

\section{Erf model for the intrinsic density profile}

\label{sec:erf}

In the previous section we have obtained general expressions  that
rely on the assumption of a model of normal translations of the mean field
density profile, \Eq{ansatzv}. In order to obtain more explicit expressions for
the surface tension and the spectrum of fluctuations, it is now required
to specify  the intrinsic density profile.

The precise dependence of $\rho_{\pi}(z;\ellav)$ on $z$ and $\ellav$ is
dictated by the molecular model and the details of
the substrate. However, quite generally, we expect that for thick
adsorbed films sufficiently far from the substrate, the $z$ dependence
in the neighborhood of $z=\ellav$ becomes independent of $\ellav$. In this
limit, we can hope to obtain general expressions that will not depend
on precise details of the substrate.

As suggested previously,\cite{macdowell13,macdowell14,benet14b}  
we consider intrinsic density profiles which satisfy the
following constraint:
\begin{equation}\label{eq:erfap}
 ( z - \ellav) \frac{d \rhoi}{d z} \approx - \xi_e^2  \frac{d^2 \rhoi}{d z^2}  
\end{equation} 
where $\xi_e$ is a phenomenological length scale of the order of the correlation
length. It is expected that this approximation is generally
 exact up to first order for free liquid-vapor interfaces, 
provided the location of the interface is
chosen at the point of $\rhoi(z)$ with maximum slope. Particularly, the
approximation is exact for a model density profile with the shape of an error
function. For this reason, we will call this the {\em Erf approximation}.

\subsection{Film height dependent surface tension}

\label{subsec:erf-1}

As summarized in section \ref{subsec:summary}, 
the surface tension of the adsorbed film is 
given by \Eq{ggammaq0}. 
The first contribution, $\gamma_0(\ellav)$,
 is dictated by the distorted liquid-vapor
density profile only (i.e., \Eq{gammasgtv1} or \Eq{gamma0vdw}) and does not 
explicitly depend
on the substrate properties. In the Erf approximation, the liquid-vapor
density profile has attained already its asymptotic shape, so that
$d\rhoi/d\ellav$ is equal to $d\rhoi/dz$, and therefore $\gamma_0(\ellav)$ is
essentially constant and equal to $\gamma_{lv}$. The only dependence 
on $\ellav$ arises from the truncation of the Gaussian tail of $d\rhoi/dz$ by
the lower bound of the integrals in either \Eq{gammasgtv1} or \Eq{gamma0vdw}. 
Obviously, such effect is negligible for $\ellav \gg \xi_e$. For smaller
$\ellav$, solving the integral explicitly
would give an $Erf$ function for $\gamma_0(\ellav)$.
However, considering the crudeness of the model, taking this result as
a quantitative statement is not warranted. Only the fact that the
 $\ellav$ dependence is in the range of the bulk correlation length 
is to be trusted, in agreement with results for the more elaborate
double parabola model in the Fisher-Jin theory,\cite{jin93} and the
Non-Local theory.\cite{parry06}

The second contribution to the surface tension, $\Delta \gamma(\ellav)$, 
results from the influence of the external field on the liquid-vapor
interface. By plugging the Erf approximation, \Eq{erfap} into \Eq{gammafilmv},
we obtain:\cite{macdowell13,macdowell14,benet14b}
\begin{equation}\label{eq:gammag}
  \Delta\gamma(\ellav) = \xi_e^2 g_{ext}''(\ellav)
\end{equation} 
where $g_{ext}''(\ellav)$ is the second derivative of the external 
contribution to the interface potential 
\begin{equation}\label{eq:gpripri}
  g_{ext}''(\ellav) = \int V(z) \frac{d^2 \rhoi(z;\ellav)}{d \ellav^2} dz
\end{equation} 
Notice that in the language of colloidal science, $g''$ corresponds to
minus the derivative of the {\em disjoining pressure}. In this way,
it is possible to relate the $\ellav$ dependence of $\Delta\gamma(\ellav)$ to
a measurable experimental property. Also note
\Eq{gammag} is consistent with
predictions from the Non-Local theory of interfaces.\cite{bernardino09}

\begin{table}
\begin{tabular}{l|ccccccccc}
\hline
substrate/fluid/vapour & $A_w$~zJ & & $\frac{|A_w|}{\gamma_{lv}}$~nm$^2$ & &
$\frac{|A_w|}{A_l}$ & &
$\ellav_{1/8}$~nm & & $\ellav_{1/8}$~nm \\
\hline
quartz/water/air & -8.7 & & 0.12 & & * & & 1.1 & & * \\
quartz/octane/air & -7.0 & & 0.32 & & 0.16 & & 1.4 & & 1.3 \\
rutile/water/air$^b$ & -98 & & 1.4 & & * & & 2.0 & & * \\
$\alpha$-alumina/octane/air$^a$ & -47.5 & & 2.2 & & 1.1 & & 2.2 & & 2.1 \\
rutile/octane/air & -94 & & 4.3  & & 2.1 & & 2.6 & & 2.4  \\
CaF$_2$/Liq-Helium/vapour & -5.9 & & 49. & & 10.3 & & 4.8 & & 3.6 \\
\end{tabular}
\caption{\label{tab_range} Table of surface properties for selected substrate/fluid pairs
and order of magnitude estimates of $\ellav_{1/8}$.
Data for $A_w$ from Ref.\onlinecite{israelachvili91}, except $^a$,
from Ref.\onlinecite{blake75} and $^b$, synthetic data from
Ref.\onlinecite{israelachvili91}. Rest of entries obtained using
$A_l=45.$~zJ and $\gamma_{lv}=21.6$~mNm$^{-2}$ for n-octane at 292~K; 
$\gamma_{lv}=73$~mNm$^{-2}$
for water at 292~K, and
$A_l=0.57$~zJ  and $\gamma_l=0.12$~mNm$^{-2}$ for He at
4~K.\cite{israelachvili91}
$\ellav_{1/8}$ obtained from \Eq{l18eq1} (5th column) 
and \Eq{l18eq3} (6th column), assuming
$\xi_e=\dhs=0.35$~nm. Note: 1~zJ=10$^{-21}$~J and 1~mNm$^{-2}$=1~zJnm$^{-2}$. }
\end{table}
Whence, for wall-fluid interactions with a range larger than the bulk
correlation length, we expect that the zero wave-vector dependent
surface tension will obey:
\begin{equation}\label{eq:gerfc}
   \gamma(\ellav) = \gamma_{lv} + \xi_e^2 g_{ext}''(\ellav)
\end{equation} 

For stable or partially stable wetting films, $g_{ext}''$ is always positive, so
that typically for thick films it is expected that $\gamma(\ellav) >
\gamma_{lv}$. This predictions has been confirmed recently for two different
models of short-range fluids in the presence of an algebraically decaying 
external field.\cite{macdowell13,benet14b}

For real fluids, exhibiting long range fluid-fluid interactions, the interface
potential is usually characterized in terms of the Hamaker constant, $A_w$,
as:
\begin{equation}
   g_{ext}(\ell) = - \frac{A_w}{12\pi\ell^2}
\end{equation} 
with $A_w<0$ for either stable or metastable wetting films.
Accordingly, we can write:
\begin{equation}
   \gamma(\ellav) = \gamma_{lv} + \xi_e^2 \frac{|A_w|}{2\pi\ellav^4} 
\end{equation} 
Clearly, $\gamma(\ellav)$ falls steeply to its asymptotic value, but could
increase much for sufficiently thin wetting films.

To asses the length-scale where the film height dependence of the surface
tension is significant, we define $\ellav_{1/8}$ as that film height
resulting in a 12.5\% increment of $\gamma(\ellav)$. Accordingly,
we find:
\begin{equation}\label{eq:l18eq1}
 \frac{\ellav_{1/8}}{\xi_e} = \left ( 
       \frac{4|A_w|}{\pi\gamma_{lv}\xi_e^2}
\right )^{1/4}
\end{equation} 
where $\ellav_{1/8}$ is expressed in units of $\xi_e$, since it is not
meaningful to describe a film of thickness smaller than the interface width.

In order to asses $\ellav_{1/8}$, we need
simple estimates for $A_w$ and $\gamma_{lv}$. 
Dietrich and Schick considered the general problem of fluid adsorption
on a substrate for systems dominated by long range dispersive
forces. They obtained expressions for the surface tension and
Hamaker constants in terms of integrals over pair potentials.\cite{dietrich86} 
In order to exploit those results, we consider  a simple model
with pair interactions made of a hard sphere repulsive interaction
of diameter $\dhs$, and a dispersion term 
$-\epsilon \dhs^6/r^6$ (Sutherland potential). 
Using integrals for the $r^{-6}$ dispersion tail
borrowed from Ref.\onlinecite{gregorio12}, it is possible
to quantify the results of Ref.\onlinecite{dietrich86} for
$A_w$ and $\gamma_{lv}$ (Appendix D). Replacing the
corresponding expressions in \Eq{l18eq1}, we obtain:
\begin{equation}\label{eq:l18eq2}
  \frac{\ellav_{1/8}}{\xi_e} = 2 \sqrt{\frac{\dhs}{\xi_e}} 
   \left (  \frac{\epsilon_w\dhs_w^6 \rho_w-\epsilon\dhs^6\rho_l}
            { \epsilon\dhs^6 (\rho_l - \rho_v)} \right )^{1/4}
\end{equation} 
where $\epsilon_w$ and $\dhs_w$ are  energy and range
parameters for the substrate-fluid pair potential, while $\rho_w$ is
the substrate's number density.

At high temperatures, close to the adsorbate's critical point, the
term in parenthesis increases slowly, but since $\xi_e$ scales as the 
correlation
length, the prefactor $\dhs/\xi_e$ decreases at a faster rate. As a result,
$\ellav_{1/8}/\xi_e$ vanishes close to the critical point.

For temperatures well below the critical point of the adsorbed fluid,
$\xi_e\approx\dhs$, while $\rho_l\gg \rho_v$. As a result, it is
possible to relate the term inside the parenthesis with a ratio
of Hamaker constants (Appendix D):
\begin{equation}\label{eq:l18eq3}
 \frac{\ellav_{1/8}}{\xi_e} = 2 \left ( 
      \frac{|A_w|}{A_l}
\right )^{1/4}
\end{equation} 
with  $A_l$ the Hamaker constant of two liquid slabs
interacting across vacuum.  The ratio $A_w/A_l$ typically falls in the
range $10^{-1}-10^1$, so that the length-scale where
 $\gamma(\ellav)$ differs significantly from $\gamma_{lv}$ is not larger than 
a few  interface widths (Table \ref{tab_range}). 
In fact, under the assumptions mentioned
at the beginning of the paragraph, the ratio $A_w/A_l$ is very nearly equal
to the spreading coefficient (Appendix D). Accordingly,  we expect 
$\ellav_{1/8}/\xi_e$ to be larger for substrate/fluid pairs above the wetting 
temperature.

\begin{figure}
\includegraphics[width=0.7\textwidth]{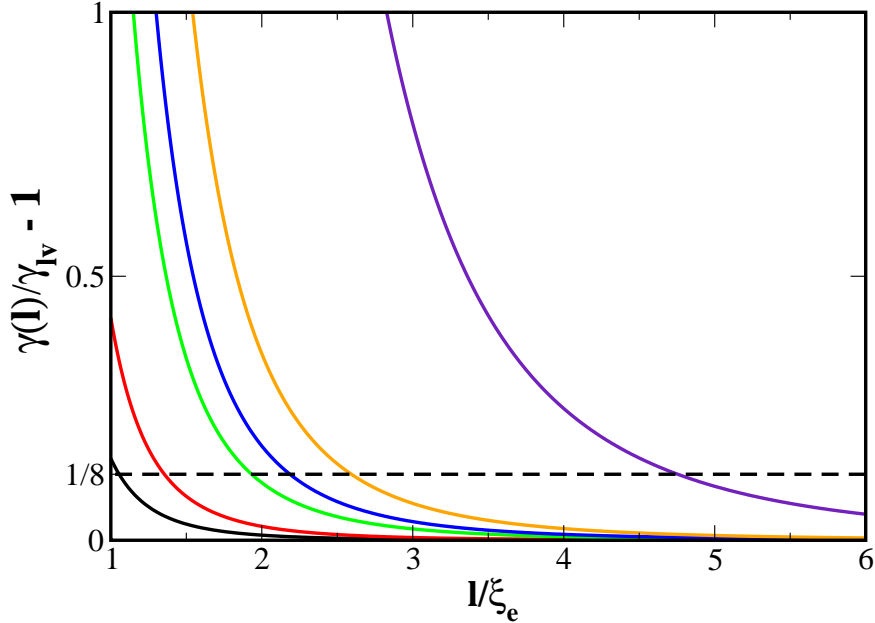}
\caption{\label{fig:gamma_pairs} Plot of $\gamma(\ellav)$ for a number
of substrate/fluid pairs. From left to right
water on quartz (black), octane on quartz (red), water on rutile (green),
octane on $\alpha$-alumina (blue), octane on rutile (orange) and
Liquid Helium on CaF$_2$ (indigo). The dashed horizontal line
indicates a 1/8 increment over the asymptotic surface tension $\gamma_{lv}$.
}
\end{figure}

Figure \ref{fig:gamma_pairs} displays $\gamma(\ellav)$ as a function of $\ellav$
for a number of different fluid/substrate pairs with $\ellav_{1/8}$ ranging
from about 1 to 5 times $\xi_e$. Clearly, the effect of the disjoining
pressure on $\gamma(\ellav)$ decays very fast, but can yield surface tensions
several times larger than $\gamma_{lv}$ for systems exhibiting a large
ratio of Hamaker constants $|A_w|/A_l$, such as the pair rutile/octane/air
and CaF$_2$/Liquid Helium/vapour.

\subsection{Capillary wave broadening}

Using \Eq{erfap} in either \Eq{avrhot} or \Eq{convapp}, 
we obtain for the thermally averaged
density profile the following result:
\begin{equation}
  \prom{\rho(\rvec{}{};\funcL)}{\Xi} = \rhoi(z;\ellav) +
 \frac{1}{2}  \frac{d^2\rhoi(z;\ellav)}{dz^2} \Delta_{cw}^2
\end{equation} 
where $\Delta_{cw}^2$ dictates the amplitude of capillary wave broadening of
the intrinsic density profile. Here, it is given as the sum of two
different contributions:
\begin{equation}\label{eq:fulldelta}
  \Delta_{cw}^2 = \Delta_{0}^2 + \Delta_{1}^2
\end{equation} 
The first one corresponds to the broadening due to mere translation of
the profile, and corresponds to
the result of classical capillary wave theory:
\begin{equation}\label{eq:delta0r}
   \Delta_{0}^2 = \prom{\delta\film^2}{\Xi}
\end{equation} 
The second one stems from distortions of the profile due to
the finite gradient of
interface fluctuations,\cite{macdowell13,macdowell14}  
and unavoidably
mixes intrinsic contributions (as dictated by $\xi_e$), and capillary
wave distortions  (as implied by the fluctuations of the film gradient):
\begin{equation}\label{eq:delta1r}
   \Delta_{1}^2 = \xi_e^2 \prom{(\nabla_{\rpar}\film)^2}{\Xi}
\end{equation} 

The intensity of specular reflectivity measurements consistent with
the above results may be obtained by replacing \Eq{erfap} into \Eq{specular}:
\begin{equation}
 I_{\rm spec}(Q_z)   =   \rhoi(Q_z)^2 
 \left [ 1 + 
       Q_z^2 \Delta_{cw}^2
 \right ]
\end{equation} 
where $\rhoi(Q_z)$ is the Fourier transform of $\rhoi(z)$, while
$\Delta_{cw}^2$ is now given by \Eq{fulldelta}, with:
\begin{equation}\label{eq:delta0f}
  \Delta_{0}^2 = \sum_{\qvec} \prom{\film^2(q)}{\Xi}
\end{equation}
and 
\begin{equation}\label{eq:delta1f}
   \Delta_{1}^2 = \xi_e^2 \sum_{\qvec} q^2 \prom{\film^2(q)}{\Xi}
\end{equation} 
Because of Parseval's theorem, the results \Eq{delta0r} and \Eq{delta0f} 
for $\Delta_{0}^2$, as well as 
 \Eq{delta1r} and  \Eq{delta1f} for $\Delta_{1}^2$ are equivalent.

In order to obtain explicit results for $\Delta_{cw}^2$, we approximate
the sum of Fourier components in \Eq{delta0f} and \Eq{delta1f} to
an integral, i.e., $\sum_{\qvec} \to \frac{A}{4\pi^2}\int d\qvec$, and
use \Eq{ncws} for the spectrum of surface fluctuations, whence:
\begin{equation}
  \Delta_{cw}^2 = \frac{k_BT}{2\pi} \int_{q_{min}}^{q_{max}}
            \frac{1+\xi_e^2 q^2}{g''(\ellav)+\gamma(\ellav;q) q^2}
          d\qvec
\end{equation} 
where $q_{min}=2\pi/L$ is the lowest possible wave-vector consistent with the
system's lateral size, as dictated by $L$, while $q_{max}$ is an upper
wave-vector cutoff.
A closed expression for the general case of a fluid 
with short and long range forces (i.e., finite $\mu$) is not possible.
Fortunately, recent studies suggest that the contribution of the singular
term $q\ln q$ in $\gamma(\ellav;q)$ is very small, so that most likely
it is possible to describe $\Delta_{cw}^2$ assuming $\mu=0$.\cite{chacon14} 
Also, notice the requirement of a finite interface width implies
$\kappa(\ellav)$ is a  positive coefficient.\cite{chacon05,tarazona07,hofling15}
In that case,
the integral may be solved analytically and approximated with good
accuracy to the following result (Appendix E):
\begin{equation}\label{eq:ncwrg1}
\begin{array}{lll}
  \Delta_{cw}^2 & = & \displaystyle \frac{k_B T}{4\pi\gamma(\ellav)} \left [
  \frac{ \xi^2_{\parallel} - \xi^2_{e} }{\xi^2_{\parallel} - 2\xi^2_{\kappa}} 
\right ] \ln \left (
  \frac{ 1 + \xi_{\parallel}^2 q_{\rm max}^2}{1 + \xi_{\parallel}^2 q_{\rm
min}^2 } \right ) 
\\ & & \\ & + & 
\displaystyle   \frac{k_B T}{4\pi\kappa(\ellav)}
\left [   \frac{\xi_{\parallel}^2\xi^2_{e}   - 
  ( \xi^2_{e}  + \xi_{\parallel}^2 ) \xi^2_{\kappa}
   }{ \xi_{\parallel}^2  - 2 \xi^2_{\kappa}  } \right ] \ln \left (
   \frac{ \xi_{\parallel}^2  - ( 1 - \xi_{\parallel}^2\, q_{\rm max}^2 )
\xi^2_{\kappa}  }
        { \xi_{\parallel}^2  - ( 1 - \xi_{\parallel}^2\, q_{\rm min}^2 )
\xi^2_{\kappa}  } \right )
\end{array}
\end{equation}
where $\xi_{\parallel}^2=\frac{\gamma(\ellav)}{g''(\ellav)}$ plays the 
role of a 
parallel correlation length for interface fluctuations and
$\xi_{\kappa}^2=\frac{\kappa(\ellav)}{\gamma(\ellav)}$ may be interpreted as
the length-scale below which bending the interface becomes too expensive.
Notice that the contributions of gradient fluctuations in the interface 
roughening (\Eq{delta1r} or \Eq{delta1f}),
may be readily recognized as those terms linear in $\xi_e^2$.

In the limit where both $\xi_e^2$ and $\xi_{\kappa}^2$ are allowed to vanish,
\Eq{ncwrg1} recovers the result of classical capillary wave theory, albeit
with a film height dependent surface tension. Relaxing
the constraint $\xi_{\kappa}^2=0$ while keeping $\xi_e^2=0$, \Eq{ncwrg1} becomes
an extended capillary wave theory that naturally provides
an upper wave-vector cutoff   $q_{\rm max}^2=\xi_{\kappa}^{-2}$.
Taking into account the fluctuations of the film gradient requires
to relax the constraint  $\xi_e^2=0$, but in this case the
bending rigidity coefficient $\kappa$ is not sufficient to
provide for an ultraviolet cutoff.

In order to find plausible values for the unknown parameters $q_{max}$,
and $\xi^2_{\kappa}$, in terms of $\xi_e$, it seems natural to
consider the result for $\Delta_{cw}^2$ in the limit of
vanishing external field ($\xi_{\parallel}^2\to\infty$):
\begin{equation}\label{eq:cwnf}
\Delta_{cw}^2 = 
\frac{k_B T}{2\pi \gamma_{lv}} \ln\left ( \frac{q_{max}}{q_{min}} \right )
- \frac{k_B T}{4\pi \gamma_{lv}} \left [ 1 - \frac{\xi_e^2}{\xi_{\kappa}^2}
  \right ]
 \ln( 1 + q_{max}^2 \xi_{\kappa}^2 )
\end{equation} 
This result may be now compared with the expectations for the
capillary wave broadening from the one-loop approximation, which
holds precisely in that limit:\cite{kopf08}
\begin{equation}\label{eq:cwnf1l}
 \Delta_{cw}^2 = \frac{k_B T}{2\pi \gamma_{lv}} \ln\left (
\frac{2\pi}{q_{min}\xi_R} \right )
- \frac{k_B T}{4\pi \gamma_{lv}} \left [ 2\alpha - \frac{\pi^2}{2}\ln 3
  e^{-13/12} \right ]
\end{equation} 
Since  $\xi_R$ and $\xi_e$ describe the interface width of the 
intrinsic profile, we set $\xi_R=\xi_e$. It is then natural to
equate \Eq{cwnf} with \Eq{cwnf1l} and to
identify $\ln(q_{max}/q_{min})$ in the first expression with
$\ln(2\pi/q_{min}\xi_R)$  in the second.   This then  yields
readily $q_{max}\approx 2\pi/\xi_e$ for the wave-vector cutoff and provides 
for the bending rigidity $\kappa\approx 4\gamma_{lv}\xi_e^2$ as the solution
of a transcendental equation.

\begin{figure}
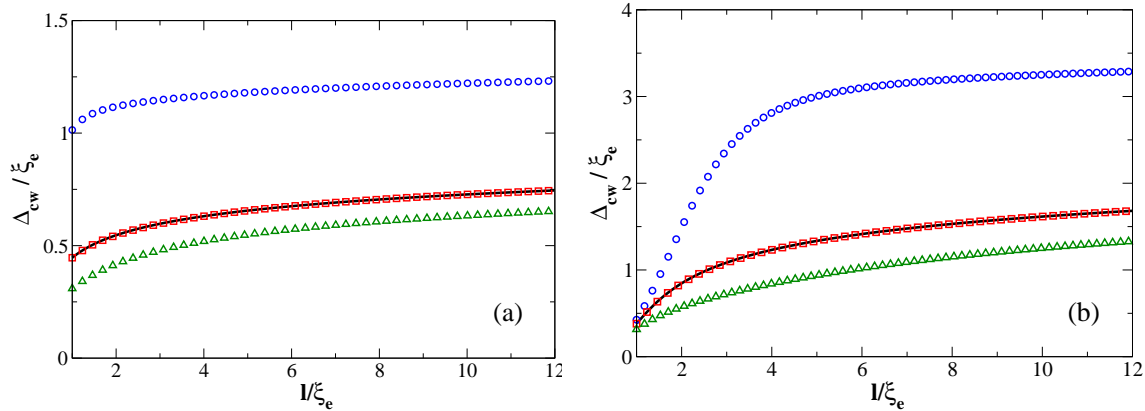

\begin{tabular}{ccc}
\includegraphics[width=0.45\textwidth]{cwb_quartz-H2O.eps} & &
\includegraphics[width=0.45\textwidth]{cwb_CaF2-He.eps} 
\end{tabular}
\caption{\label{fig:cwb}  Plot of $\Delta_{cw}$ as given by
\Eq{ncwrg1s} (symbols), compared to the classical theory (lines)
for two substrate/fluid pairs (a), water adsorbed on quartz and
(b), Liquid Helium adsorbed on CaF$_2$). 
Results are obtained for fixed $q_{max}=2\pi/\xi_e$, and
a choice of bending rigidities, corresponding to 
$\xi_{\kappa}=\xi_e/2\pi$ (blue circles), 
$\xi_{\kappa}=\xi_e$ (red squares) and
$\xi_{\kappa}=\sqrt{40/11}\xi_e$ (green triangles).
The latter choice is suggested by the one-loop approximation.
Notice that for intermediate film heights this choice requires
evaluating $\Delta_{cw}^2$ with the exact result in complex algebra 
(see Appendix E).  In all cases, $\xi_e=0.35$~nm.
%
}
\end{figure}

Taking now the limit of large system sizes, $\xi_{\parallel}^2 q_{min}^2 \ll 1$,
while allowing for a finite external field, which will usually be the relevant
experimental situation, we find for the capillary wave broadening:
\begin{equation}\label{eq:ncwrg1s}
 \Delta_{cw}^2 = \frac{k_B T}{4\pi\gamma(\ellav)}
    \left [
  \frac{ \xi^2_{\parallel} - \xi^2_{e} }{\xi^2_{\parallel} } \right ]  
          \ln \left (  \xi_{\parallel}^2 q_{max}^2  \right )
 +
 \frac{k_B T}{4\pi\gamma(\ellav)}
\left [
 \frac{ \xi^2_{e} }{\xi_{\kappa}^2} - 1 \right ] 
       \ln \left ( 1 + q_{max}^2 \xi_{\kappa}^2 \right ) 
\end{equation} 
We test this equation for strong to moderate external fields, by setting
$q_{max}=2\pi/\xi_e$ as suggested above, while allowing for a choice of
bending rigidities (Fig. \ref{fig:cwb}).
In the limit of very small external fields, 
$\xi^2_{\parallel}\to \infty$,
our result becomes equal to that of classical capillary wave theory,
except for an additive constant. However,
in the presence of a tunable external field, classical
theory predicts a broadening that is linear in $\ln\xi_{\parallel}^2$,
while our theory of normal interface translations suggests
the prefactor of the logarithmic term also depends
on the external field. 

In practice, the difference between \Eq{ncwrg1s} and the classical result
(which is recovered simply by setting $\xi_e=0$)
is mainly dictated by the second term in the right hand side of \Eq{ncwrg1s}.
If the ratio $\xi_e/\xi_{\kappa}$ differs from unity, it provides a nearly constant
shift of the capillary wave broadening that may be either
positive ($\xi_{\kappa}<\xi_e$) or negative ($\xi_{\kappa}>\xi_e$) 
and should be possible to distinguish experimentally (Fig. \ref{fig:cwb}). 

If, on the other hand  $\xi_e/\xi_{\kappa}\to 1$, the shift vanishes altogether. In
that case, the logarithmic contribution from \Eq{ncwrg1s} is hardly
distinguishable from the classical result (Fig. \ref{fig:cwb}).

In practice,  $\xi_{\kappa}$ must be considered an empirical parameter,
so that we cannot tell a priori the extent to which our result differs from
the classical theory.
By performing x-ray reflectivity experiments, it should be possible in
principle to measure $\Delta_{cw}^2$ and confirm the expectations of
\Eq{ncwrg1} and \Eq{ncwrg1s} and to provide an estimate for $\xi_{\kappa}$.
Interestingly, several x-ray diffraction experiments performed on
fluid surfaces report the need to account for a constant shift
on the results for $\Delta_{cw}$ which would be consistent with
the expectations from \Eq{ncwrg1s}  assuming
$\xi_{\kappa}<\xi_e$.\cite{ocko94,heilmann01,plech02} Unfortunately,
it is not possible to distinguish whether this shift stems from the
intrinsic width of the interface or from gradient fluctuations to the
capillary wave broadening.


As a final remark, we note that, whereas the result of \Eq{cwnf} for
$\Delta_{cw}^2$ is consistent with
the result of the one-loop approximation, \Eq{cwnf1l}, 
a stringent comparison of the individual components as
implied in \Eq{fulldelta} does not seem to match so consistently.

Indeed, from \Eq{delta0r}--\Eq{delta1r} and \Eq{ncwrg1}, in the limit of vanishing external fields, we find:
\begin{equation}\label{eq:cosa10}
\prom{\delta\film^2}{\Xi} = \frac{k_B T}{2\pi \gamma_{lv}} 
 \ln \left( \frac{q_{max}}{q_{min}( 1 + q_{max}^2 \xi_{\kappa}^2 )^{1/2}} \right
)
\end{equation} 
\begin{equation}\label{eq:cosa11}
\prom{(\nabla_{\rpar}\film)^2}{\Xi} = \frac{k_B T}{4\pi \gamma_{lv}} 
 \frac{\ln ( 1 + q_{max}^2 \xi_{\kappa}^2 )}{\xi_{\kappa}^2}
\end{equation} 
On the contrary, the comparison of \Eq{avrhot} with the one-loop result
of \Eq{rgt-tuneada} suggests the fluctuations should be, rather:
\begin{equation}\label{eq:cosa20}
\prom{\delta\film^2}{\Xi} = \frac{k_B T}{2\pi \gamma_{lv}} 
 \ln\left ( \frac{2\pi \,e^{-\alpha}}{q_{min}\xi_R} \right)
\end{equation} 
\begin{equation}\label{eq:cosa21}
\prom{(\nabla_{\rpar}\film)^2}{\Xi} = \frac{k_B T}{4\pi \gamma_{lv}} 
\frac{3}{2} \frac{\ln ( 3\, e^{-\frac{13}{12}} )}{\xi_{R}^2}
\end{equation} 

Matching \Eq{cosa10} with \Eq{cosa20} and \Eq{cosa11} with \Eq{cosa21} provides a system of 
two equations with two unknowns, $q_{\max}$ and $\xi_{\kappa}$,
but unfortunately, the only solution yields
the result $q_{max}^{-2}=244\xi_e^2$ and $\kappa=-243\gamma_{lv}\xi_e^2$.
The origin of the unexpected small cutoff and negative $\kappa$ lies
in the result for $\prom{(\nabla\film)^2}{\Xi}$ in \Eq{cosa21}, which is
close to zero (since $\ln 3 -13/12\approx 0$) and can only match
\Eq{cosa11} if we accept a negative $\kappa$.

The difference of this unsatisfactory comparison with that performed
previously, which provided results for $q_{max}$ and $\kappa$ closer to
expectations is whether one interprets the term $-\frac{k_BT}{2\pi\gamma}\alpha$
in \Eq{cwnf1l} as
belonging to either $\prom{(\delta\film)^2}{\Xi}$  or
$\prom{(\nabla\film)^2}{\Xi}$. In view of this discussion, the latter
interpretation seems more justified.


\section{Comparison with exact results}

\label{sec:exact}

Before closing, we test our results with  an exact solution of
the Landau-Ginzburg-Wilson Hamiltonian under an external field.
A solution of this system for arbitrary external fields $V(z)$,
is generally not possible. However, in an exceptional and somewhat
forgotten paper, Zittartz noticed many years ago that  this problem
may be remedied for an external field of $\tanh(z)$ form.\cite{zittartz67}

Particularly, Zittartz considered the free energy functional
\Eq{sgtv} in the lattice gas analogue,  with the usual
biquadratic bulk free energy $f(\rho)=\alfa \rho^4 - \exilon \rho^2$
and an external field:
\begin{equation}\label{eq:vzitta}
   V(z) = 2 \strength \left ( \frac{\exilon+\strength}{2\alfa} \right )^{1/2} 
               \tanh\left (
                        \frac{ 1 }{2}  \frac{z-\ellav}{\xi_u}
                  \right )
\end{equation}
where
\begin{equation}
     \xi_u = \frac{1}{2} \left ( \frac{\exilon+\strength}{C} \right )^{-1/2}
\end{equation}
This external field is unusual, because it has
its origin at the interface position. Accordingly,
the free energy depends only on the field strength $\strength$,
and not on the interface  position.

The exact mean field (intrinsic) density profile is:\cite{zittartz67}
\begin{equation}\label{eq:mzitta}
  \rhoi(z) =  \left ( \frac{\exilon+\strength}{2\alfa} \right )^{1/2}
               \tanh\left ( 
                         \frac{ 1 }{2}  \frac{z-\ellav}{\xi_u}  
                  \right )
\end{equation} 
Notice that the role of $V(z)$ is to pin exactly the interface
at $z=\ellav$ and set the interface width $\xi_{\strength}$. 

Armed with this solution, we can now assess several of the results
of section \ref{cap:srfef}, \ref{subsec:summary} and \ref{subsec:erf-1}.

First consider the surface tension as predicted by the
Fisher-Jin theory for a system with short range forces in an external
field $V_0(z)$ equal to $V(z)$ above. 
Using \Eq{gammasgtv1}, with \Eq{mzitta}
for the density, we obtain in closed form:
\begin{equation}\label{eq:gammazitta}
 \gamma_{\strength} = \frac{2}{3} 
     \frac{C^{1/2}}{\alfa}(\exilon + \strength)^{3/2} 
\end{equation} 
where we have added the subindex $\strength$ next to $\gamma_{\strength}$
 in order to stress
the explicit dependece on the external field that we have assumed.

Clearly, as $\strength\to 0$, $\gamma_{\strength}$ splits into
$\gamma_0 = \frac{2}{3} \frac{C^{1/2}}{\alfa}\exilon^{3/2}$, for
the surface tension in zero field, and $\Delta \gamma =
\frac{ (\exilon C)^{1/2} }{\alfa}\strength$
for linear corrections in the field
strength.

Now consider the perturbative result, \Eq{gammafilmv}, for the correction
of $\gamma_0$ due to the external field $V(z)$, which, 
using again \Eq{mzitta} for the density yields:
\begin{equation}
   \Delta \gamma = \frac{C^{1/2}(\exilon + \strength)^{1/2} }{\alfa}\strength
\end{equation} 
Clearly, in the limit $\strength\to 0$, this result
 provides exactly the same leading order correction
to $\gamma_{\strength}$ that was obtained using $\Eq{gammasgtv1}$
in the paragraph above. 
This attests to the consistency of our approach. Particularly,
it shows that the approximation used in \Eq{gammafilmv} remains very
robust, even though \Eq{rhoexpds} does not yield the exact limit
of density decay at infinity. 

Now, consider the calculation of $g''$, which can be performed by
plugin the density profile of \Eq{mzitta} into \Eq{gpripri}. Again, the result
may be obtained in closed form as:
\begin{equation}\label{eq:gzitta}
   g''_{ext} = \frac{4}{3} \frac{(\exilon + \strength)^{3/2}}{\alfa
C^{1/2}} \strength
\end{equation} 
using the result for the bulk correlation length in zero field,
$\xi_0^2= \frac{1}{4} C/\exilon$, together with 
$\Delta \gamma = g'' \xi_0^2$ (c.f. \Eq{gammag}) we find,
to linear order in the field strength:
\begin{equation}
      \Delta \gamma = \frac{1}{3} \frac{ ( C \exilon )^{1/2} }{\alfa} \strength
\end{equation} 
Whence, the approximate solution \Eq{gammag}, provides also
the correct result, with an empirical measure of the interface width
$\xi_e=\sqrt{3}\xi_0$. This is a very handy result, because most
 often neither the density profile nor the external field
are known. Therefore,
the explcit results \Eq{gammasgtv1} or \Eq{gammafilmv} are not
practical. On the contrary, the first derivative $g'$ is the negative of the
disjoining pressure  and can be measured experimentally.

So far we have tested that the alternative results
 \Eq{gammasgtv1}, \Eq{gammafilmv} 
and \Eq{gerfc} for the film height dependent surface
tension are  consistent.  But it remains to show that
these corrections to the surface tension  have their signature stamped
in the spectrum of surface fluctuations, as suggested in \Eq{ncws}.

To show this, consider the density-density correlation function
predicted by capillary wave theory, \Eq{diffuse}:
\begin{equation}\label{eq:paircw}
      G_{cw}(z_{1},z_{2}{};q) =
\frac{d\rhoi(z_1)}{d\ell} \frac{d\rhoi(z_2)}{d\ell} \left\langle
 \film^2(q)\right\rangle 
\end{equation} 
with $\left\langle\film^2(q)\right\rangle$ given by \Eq{ncws}.

This result may be compared with 
the correlation function of the Landau-Ginzburg-Wilson Hamiltonian
as discussed by Zittartz and Jasnow.\cite{zittartz67,jasnow84} Exact solutions
exist in closed form.\cite{zittartz67} However, for this
system it is more convenient to exploit the fact that 
$G(\rvec{1}{},\rvec{2}{})$ is a Green's function.
Accordingly, it may be expressed as an eigenvalue expansion as follows:
\begin{equation}
  G(z_{1},z_{2}{};q) = \sum_n \frac{ \phi^{*}_n(z_1) \phi^{}_n(z_2)
}{\lambda_n}
\end{equation} 
where $\phi_n(z)$ and $\lambda_n$ are the solutions of
the eigenvalue equation:
\begin{equation}
\left [  -C \frac{d^2}{d z^2} + C q^2 - 2\exilon +12\alfa \rhoi^2(z) 
  \right ]\phi(z) = \lambda \phi(z)
\end{equation} 
In the quantum mechanical analogy, with $\rhoi(z)$ of hyperbolic tangent form,
this is the Schr\"odinger equation for a shifted P\"oschl-Teller potential,
whose exact solutions are well known.\cite{lekner07}

The first two eigenvalues of the  P\"oschl-Teller well 
correspond to states  bound
to the potential, which are naturally related to purely interfacial 
contributions to the correlation function. The remaining eigenvalues
lay in the continuum and may be considered as corresponding to bulk 
correlations perturbed by the interface. 

The bound state  of lowest energy is a soft mode which
merely describes the displacement of the interface, without change
of the density profile.\cite{zittartz67,jasnow84} Its eigenfunction is
$\phi_1(z) = d\rhoi/dz$, and the corresponding eigenvalue is:
\begin{equation}
   \lambda_1 = 2\strength + C q^2
\end{equation} 
Clearly, in the limit of $\strength\to 0$, $\lambda_1\propto q^2$,
and we can therefore identify this mode as the translation mode
of the capillary wave Hamiltonian. As the field is switched on,
the first eigenvalue merely describes how the translational
mode is modified by the external field.

From \Eq{gammazitta} and \Eq{gzitta}, one readily finds that the 
ratio $\frac{1}{2} C/\strength$ is precisely the ratio
of $\gamma_{\strength}$ to $g''$ in the Zittartz model. 
Accordingly, it follows that, under the external field,
\Eq{vzitta},
the translational mode of the correlation function is:
\begin{equation}\label{eq:pairtras}
      G_{\rm tras}(z_{1},z_{2}{};q) \propto
\frac{d\rhoi(z_1)}{d\ell} \frac{d\rhoi(z_2)}{d\ell} 
  \frac{k_B T} { g'' + \gamma_{\strength} q^2 }
\end{equation}
where 
$\gamma_{\strength}$ is given exactly to linear
order in the field strength by either \Eq{gammasgtv1} or \Eq{gammafilmv}.

Comparison of this result  with \Eq{paircw} in the limit of 
small wave-vectors $q\ll \kappa$, indicates that our
result for the spectrum of interface fluctuations, \Eq{ncws},
is also exact to linear order in $\strength$.

This  reveals clearly the strengths and limitations of
the capillary wave approach. On the one hand,
we have showed that considering explicitely 
perpendicular rather than merely vertical translations
of the interface is sufficient to describe exactly
to linear order in the field strength the long wavelength 
surface fluctuations of the translational mode. 
On the other hand, capillary wave
theory, up to this level of detail, cannot do anything
else. i.e., the remaining surface mode, 
with explicit $q$ dependence
is completely beyond reach, and cannot be described at
all without considering explicitely perturbations of the
intrinsic density profile. Likewise,  modes in the
continuum, which can be identified with bulk correlations
perturbed by the interface, are also beyond
the level of description that can be achieved with
capillary wave theory.
Presumably, the non-local theory of interfaces shares similar
limitations, since there the corrections to the density profile
are given merely by the bulk correlation function.\cite{bernardino09}
Some of these limitations have been discussed by studying
the correlation function of the double parabola model, which,
unlike more elaborate biquadratic free energies,
has only one bound surface state.\cite{parry16} 
A promising approach to single out the surface translation mode 
from the full correlation function has been suggested 
recently.\cite{hernandez-munoz16}


\section{Conclusions}

In this paper we have considered the phenomenological extension of classical capillary wave
theory to the case were the density is dictated by the normal distance to the
local interface position (\Eq{ansatz}).  This idea seems justified on intuitive
grounds and symmetry considerations, at least for long wave-length fluctuations
in the absence of an external field. Recently, it has been shown that the
hypothesis remains accurate even for liquid films close to the three phase contact
line.\cite{nold16} Not surprisingly, the approach has been 
explored previously, starting with an apparently overlooked contribution by
Davis many years ago.\cite{davis77,diehl80,kawasaki82,stecki98,stecki01,mecke99b} However,
it would seem that some important consequences had not been recognized. Other
recent studies have rather attempted to assess the role of interface
curvature.\cite{mecke99b,blokhuis09,parry11} Such effects can  be incorporated as an effective 
wave-vector dependent surface tension, and appear as corrections of order $q^4$ in the capillary wave
spectrum. Unfortunately, it would seem that both on theoretical and experimental
grounds the study of
such corrections from the spectrum of surface fluctuations poses serious
difficulties.\cite{tarazona07,parry16} On the other hand, 
we have shown that in the presence of an external-field
the assumption of an interface profile
along lines normal to the interface results in the coupling
of surface and bulk fluctuations. This produces corrections 
of order $q^2$ which feed into the surface
tension and are linear in 
$g''_{ext}(\ellav)$.\cite{macdowell13,macdowell14,benet14b} 
Comparing our results to a more formal approach based on linear
response theory indicates that the simple phenomenological extension considered
here might be sufficient to identify the most relevant corrections to the
classical theory for flat substrates  away from the strong fluctuating
regime.\cite{bernardino09}

The first, and most immediate implication of our approach is that already to first order in
deviations from planarity, the theory picks up an additional capillary wave
broadening mechanism, with contributions that are given by the fluctuations of
the film profile gradient (\Eq{avrhot}). Such additional broadening may be captured in terms
of a `convolution' approximation, by assuming that the normal distance of a
point to the interface is Gaussian distributed (\Eq{convapp}). The effect 
on the average density profile could be measured in principle by the
specular contribution to x-ray surface scattering (\Eq{specular}).

In the absence of an external field, we show that this phenomenological theory is consistent with
renormalization group theory in the one loop approximation.\cite{jasnow78,kopf08}
The success of this comparison indicates that it is meaningful to decouple bulk and 
capillary wave fluctuations even close to the critical point, or equivalently,
that one can assume the Fisk-Widom scaling form of a density profile prior to renormalization
of capillary wave-fluctuations. This holds  provided the density is given along normals to
the film profile (\Eq{fiskwidom}). Applying this condition to a
Cahn-Hillard square gradient approximation, one recovers the capillary wave
Hamiltonian exactly 
in the limit of small curvature
(\Eq{cwhmf}).\cite{davis77,huse85}

We have further extended the theory of normal translations to the case of
adsorbed liquid films (\Eq{ansatzv}). For systems of short range forces only, this recovers
the Fisher-Jin theory of short range wetting
(\Eq{hamsrsg},\Eq{gammasgtv1}),\cite{jin93,fisher94} and introduces a film
height dependence of the surface tension which has been identified recently in
computer simulation experiments.\cite{fernandez12,fernandez15} In the presence of a long
range forces, the external field couples to the film gradient fluctuations, and
results in an explicit dependence of the surface tension on the  field
that is consistent with expectations from the Non-Local theory of interfaces
(\Eq{gammafilmv}).\cite{bernardino09}  
The signature of this coupling appears explicitely in the 
spectrum of surface fluctuations, \Eq{ncws}. Comparison with
results for an exactly solvable model of the LGW Hamiltonian in
an external field indicate that  \Eq{ncws}, together with
either \Eq{gammasgtv1} and \Eq{gammafilmv} reproduces
exactly to linear order in field strength the translational
mode of the density correlation function (c.f: section \ref{sec:exact}).
Including long-range dispersive forces, the theory yields the well known
logarithmic singularity of the wave-vector dependent surface
tension (\Eq{gammaqvdw},\Eq{ggamma}).\cite{napiorkowski93,mecke99b,blokhuis09}

We have studied a simple
model of adsorbed films which assumes the liquid-vapor density profile is independent of the
proximity to the substrate and takes the form of an error function. Under this
simplifying assumption, our approach allows to write the film height dependent
surface tension explicitly in terms of the disjoining pressure (\Eq{gerfc}). 
For wetting films, this results in a strong enhancement of the
surface tension that has  been verified in computer simulation
experiments.\cite{macdowell13,benet14b} Our qualitative calculations indicate that
the range where the film height dependent surface tension may be measured lays in
the sub-nanometer range (Table \ref{tab_range}). 
 For such thin films, the corrections to classical
theory become significant and could be measured {\em in principle} by means of x-ray scattering
experiments (\Eq{ncwrg1s}), where the specular reflectivity allows to measure the
interface width, $\Delta_{cw}$ while the diffuse scattering  probes the exponent 
$\eta\propto k_BT/\gamma$.\cite{doerr98,pershan12}  Experimental observation of a larger $\Delta_{cw}$ and a smaller 
$\eta$ than expected from the classical theory would  provide strong indications in support of our
conclusions. Unfortunately, this task requires to disentangle the purely
translational mode of surface fluctuations from additional surface and bulk
correlations, which seems difficult at present without a very accurate model for
the full inhomogeneous pair correlation function.\cite{hernandez-munoz16}
Alternatively, these corrections could become important when
attempting to extend the capillarity approximation to adsorbed films at the
nanoscale, as revealed by recent computer simulations and atomic force microscopy
experiments which indicate the need to account for an enhanced surface tension.\cite{asay06,laube15}

\begin{acknowledgments}
This paper is dedicated to the memory of No\'e G. Almarza.

 I would like to acknowledge  collaboration with E. Fern\'andez, E. Chac\'on 
and P. Tarazona for having motivated this work. 
Very helpful comments from the reviewers are also acknowledged. 
I am also indebted to J. Horbach
for drawing my attention to Ref.\onlinecite{kopf08} and A. Nold,
D. N. Sibley, B. D. Goddard and S. Kalliadasis for providing me numerical
evidence in support of \Eq{ansatzv}; Thanks also to  A. O. Parry and C. Rasc\'on
for helpful discussions.  
\end{acknowledgments}

\appendix



\section{Derivation of \Eq{specular} and \Eq{diffuse}}

In order to obtain \Eq{specular} and \Eq{diffuse}, we notice that all odd terms in $\film(\rpar)$ vanish
after thermal averaging of the density correlation function, \Eq{rhorho}. This 
means that  $\langle \delta\hper\rangle$ only
retains the  term in $(\nabla \delta\film)^2$. For this reason, we can write:
\begin{equation}
 \begin{array}{lll}
\int d \rpar_1 d \rpar_2 \langle \delta\hper(\rpar)\rangle 
   e^{i{\bf Q}_{\rpar}\cdot (\rpar_1-\rpar_2)}
 & = & 
 \frac{1}{2} \int d \rpar \, d \Delta \rpar \, \langle
  (z - \ell) \sum_{\bf q} {\bf q}\, \film({\bf q}) e^{i{\bf q}\cdot \rpar} \cdot
      \sum_{\bf q'} {\bf q'} \film({\bf q}') e^{i{\bf q}'\cdot \rpar} \rangle \,
    e^{i{\bf Q}_{\rpar}\cdot \Delta\rpar}  \\
 & & \\
 & = & 
 \frac{1}{2} (z - \ell) \sum_{\bf q} q^2 \langle \film^2({\bf q}) \rangle \; \delta({\bf Q}_{\rpar})
 \end{array}
\end{equation} 
where we have expressed $\nabla\film(\rpar)$ as an expansion in
Fourier modes, and performed the change of variables
$\Delta\rpar=\rpar_1-\rpar_2$. For the 
average $\langle \delta\hper^2\rangle$, we proceed likewise
and  write:
\begin{equation}
 \begin{array}{lll}
\int d \rpar_1 d \rpar_2 \langle \delta\hper^2(\rpar)\rangle 
   e^{i{\bf Q}_{\rpar}\cdot (\rpar_1-\rpar_2)}
 & = & 
 \frac{1}{2} \int d \rpar \, d \Delta \rpar \,
  \langle (z - \ell) \sum_{\bf q} \film({\bf q}) e^{i{\bf q}\cdot \rpar}
         \sum_{\bf q'} \film({\bf q}') e^{i{\bf q}'\cdot \rpar} \rangle
    e^{i{\bf Q}_{\rpar}\cdot \Delta\rpar}  \\
 & & \\
 & = & 
  \sum_{\bf q} \langle \film^2({\bf q}) \rangle \delta({\bf Q}_{\rpar})
 \end{array}
\end{equation} 
where all terms beyond quadratic order in the Fourier amplitudes 
$\film({\bf q})$ have been ignored.  Finally, for the crossed correlations we write:
\begin{equation}
 \begin{array}{lll}
\int d \rpar_1 d \rpar_2 \langle \delta\hper(\rpar_1)\delta\hper(\rpar_2)\rangle 
   e^{i{\bf Q}_{\rpar}\cdot (\rpar_1-\rpar_2)}
 & = & 
  \int d \rpar_1 \, d \rpar_2 \, \langle \film(\rpar_1) \film(\rpar_2) \rangle
 e^{i{\bf Q}_{\rpar}\cdot (\rpar_1-\rpar_2)}  \\
 &  & \\
 & = & \langle \film^2({\bf Q}_{\rpar}) \rangle
\end{array}
\end{equation}
Using these transforms in \Eq{born}, we obtain \Eq{specular} and \Eq{diffuse}.

\section{Derivation of \Eq{gradexpds}}

In order to arrive at \Eq{gradexpds}, we need to consider the derivatives
of $\rhoi(z;\ell)$ for an adsorbed film, which for 
symmetry reasons  depends explicitly  on two variables,
 the distance from the wall, $z$, and the film height, $\ell$.  This differs 
from the case of a free interface, where
$\rhoi$ is a function of the single variable, $z-\ell$. Also it is required
to take into account that, by virtue of \Eq{ansatzv}, 
$\rhoi(z;\ell)$ is evaluated
for a film height, $\ell=\film + \delta \hper$ which depends explicitly on
$\rpar$ and $z$. Whence:
\begin{equation}
  \nabla \rhoi(z;\ell=\film + \delta \hper) =
\left ( \frac{\partial \rhoi}{\partial z} + \frac{\partial \rhoi}{\partial \ell}
\frac{d \ell}{d z} \right ) {\bf k} + \frac{\partial \rhoi}{\partial \ell}
\nabla_{\rpar}\ell
\end{equation} 
where ${\bf k}$ is a unit vector in the $z$ direction.
To quadratic order in $\film$, the squared gradient is then given as:
\begin{equation}\label{eq:appeq1}
 \left ( \nabla \rhoi(z;\ell=\film + \delta \hper) \right )^2 =
  \left ( \frac{\partial \rhoi}{\partial z} \right )^2 +
 \frac{\partial \rhoi}{\partial z} \frac{\partial \rhoi}{\partial \ell} 
\left ( \nabla_{\rpar}\film \right )^2 + \left ( \frac{\partial \rhoi}{\partial
\ell} \right )^2  \left  ( \nabla_{\rpar}\film \right )^2
\end{equation} 
where we have employed $d\ell/d z= \frac{1}{2}(\nabla_{\rpar}\film)^2$.
Notice that all terms in the right hand side of \Eq{appeq1} are evaluated at
$\ell=\film + \delta\hper$. In order to arrive at \Eq{gradexpds}, which has the
density profiles evaluated at the film height $\ell=\film$, we 
expand  in powers of $\delta \hper$.  Since we are only interested 
in contributions of order $(\nabla_{\rpar} \film)^2$, at most, 
it suffices to expand only the first term to linear order in
$\delta \hper$, and retain the second and third terms to 
obtain \Eq{gradexpds}. 

\section{Stationarity at the boundary and alternative
derivation of \Eq{gammasgtv1}}

%
In the minimization of \Eq{sgtv}, the density at the
wall is not prescribed {\em a-priori}, but must rather be
given as a solution of the variational problem. For this
reason, \Eq{sgtext} is necessary but not sufficient condition.
To see this, we follow Ref.\onlinecite{parry06}, and 
use $\nabla\rhoi\cdot\nabla\delta\rho =
\nabla\cdot(\nabla\rhoi\delta\rho)-\nabla^2\rhoi \delta\rho$
in \Eq{sgtextint}, together with the divergence theorem.
The stationary condition then becomes: 
\begin{equation}\label{eq:extbound}
  \int  d \rvec{}{}\,\delta\rho \left \{ 
 \frac{\partial f}{\partial \rho} - C \nabla^2 \rhoi + V_0 
\right \} + 
\int dS\, \delta\rho\,  C \nabla\rhoi\cdot{\bf n}  = 0
\end{equation} 
where {\bf n} is a unit vector perpendicular to the wall, and $dS$ denotes
integration over the surface of the wall. In variational problems
where  $\rhoi$ at the surface is prescribed the surface term 
vanishes and \Eq{sgtext} is sufficient to solve the variational problem. 
Here, we can neither assume
a priori $\delta\rho=0$, nor $ \nabla\rhoi=0$ at the wall, 
and the surface term must be retained. 

In order to obtain \Eq{gammasgtv1}, we eliminate
the integral over the first square brackets of \Eq{gammasgtv0} using
\Eq{extbound} with the choice 
$\delta\rho = (z-\film)\frac{\delta\rhoi}{\delta \ell}$. 
Performing an integration by parts two of the three terms in the 
resulting integral mutually cancel each other, and we are left with 
\Eq{gammasgtv1}.

\section{Surface properties of a substrate-fluid pair interacting
via the Sutherland potential}

The Sutherland potential is:
\begin{equation}
 u = \left \{
\begin{array}{ll}
 \infty & \quad \quad r < \dhs \\
 & \\
 -C/r^6 & \quad \quad r > \dhs
\end{array}
 \right .
\end{equation} 
In what follows, we write the constant $C$ 
as $C=\dhs^6\epsilon$.

Using the results from Ref.\cite{dietrich86} and the integrals for
$-C/r^6$ from Ref.\cite{gregorio12}, we obtain, after very tedious but
straightforward manipulations:
\begin{equation}
 \gamma_{lv} = \frac{1}{4} \pi \epsilon \dhs^4 (\rho_l - \rho_v )^2  
\end{equation} 
\begin{equation}
 \gamma_{w\beta} = \frac{1}{4} \pi\epsilon_w\dhs_w^4(\rho_w-\rho_{\beta})^2 
      + \frac{5\pi}{12}\epsilon_w\dhs_w^4\rho_w\rho_{\beta} 
\end{equation}
for $\beta={l,v}$.
\begin{equation}
 A_w = \pi^2 (\rho_v - \rho_l)(\epsilon_w\dhs_w^6\rho_w - \epsilon
\dhs^6\rho_l )
\end{equation}
Defining the spreading coefficient as $S=(\gamma_{wv}-\gamma_{wl})/\gamma_{lv}$,
we obtain:
\begin{equation}
 S = \frac{ \frac{7}{6} \epsilon_w\dhs_w^4\rho_w - 
          \epsilon \dhs^4 (\rho_l + \rho_v) }
         { \epsilon \dhs^4 (\rho_l - \rho_v ) }
\end{equation} 
For $\rho_l\gg \rho_v$ and $\dhs_w \approx \dhs$, we find the
spreading coefficient becomes:
\begin{equation}
 S \approx -\frac{A_w}{A_l} 
\end{equation} 
where
\begin{equation}
 A_l = \pi^2\epsilon \dhs^4 \rho_l^2
\end{equation} 

The results of this appendix 
improve our previous estimates of the surface
properties of the Sutherland model.\cite{gregorio12} The key
point is the precise evaluation of the dispersion integrals
within perpendicular distances $|z|<\dhs$. For $\gamma_{lv}$,
for example, our results may now be cast in terms of
$A_l$ as $\gamma_{lv}= A_l/4\pi\dhs^2$, which is very
nearly equal to the empirical result 
$\gamma_{lv}= \frac{25}{24} \frac{A_l}{4\pi\dhs^2}$
advocated by Israelachvili.\cite{israelachvili91}

\section{Derivation of \Eq{ncwrg1}}

Here we solve the indefinite integral that is required to
obtain $\Delta_{cw}^2$ in \Eq{ncwrg1}.
\begin{equation}
  I =  \int
            \frac{q+\xi_e^2 q^3}{g''+\gamma q^2 +\kappa q^4}\,  d q
\end{equation} 
In principle this integral may be obtained readily in terms
of  $\arctan$ functions, but this expression involves complex numbers and
does not allow for an obvious comparison with the results of classical capillary 
wave theory ($\xi_e=0$, $\kappa=0$), which is given in terms of the logarithmic function. In order
to reveal the similarities between both results, we factor
the denominator as:
\begin{equation}
 g''+\gamma q^2 +\kappa q^4 = (a + \gamma q^2 )(r + \frac{\kappa}{\gamma}\, q^2)
\end{equation} 
under the requirement that it becomes equal to the first
factor in the right hand side in the limit $\kappa\to 0$.
Solving for $a$ and $r$, we find:
\begin{equation}
 \left \{ 
 \begin{array}{lll}
a & = & \frac{g''}{r} \\ 
r & = & \frac{1}{2}\left ( 1 + \sqrt{1-4g''\kappa/\gamma^2} \right)
  \end{array}
 \right .
\end{equation} 
Having factored the denominator, the integral may be solved by the technique of
partial fractions,\cite{kouba13} yielding:
\begin{equation}
 I = \frac{1}{2} \frac{r\gamma-\xi_e^2 g''}{r^2 \gamma^2-g''\kappa}
   \ln ( g'' + \gamma r q^2 )
  +
  \frac{1}{2} \frac{\gamma r}{\kappa} 
  \frac{\xi_e^2 \gamma r-\kappa}{r^2\gamma^2-g''\kappa}
  \ln ( \gamma r + \kappa q^2 )
\end{equation} 
\Eq{ncwrg1} is recovered from this result in the limit of
$4g''\kappa/\gamma^2\ll 1$.  In practice, we checked that 
\Eq{ncwrg1} provides a very robust 
approximation to the exact result in  physically relevant
situations considered previously.\cite{macdowell13}

\end{document}